\documentclass[letterpaper,12pt]{article}
\usepackage[utf8]{inputenc}
\usepackage{lipsum}
\usepackage[margin=0.7in]{geometry}
\usepackage{mathptmx}
\usepackage{xfrac}
\usepackage{graphicx}
\usepackage{float}
\usepackage{caption}
\usepackage{subcaption} 
\usepackage{wrapfig}
\usepackage{amsmath}
\usepackage{amssymb}
\usepackage{epstopdf}
\usepackage{setspace}
\usepackage{dirtytalk}
\usepackage{amsmath,amssymb}
\usepackage{mathtools, nccmath}
\usepackage{fancyhdr}
\usepackage{slashed}
\usepackage[numbers,sort&compress]{natbib}
\usepackage{hyperref}
\hypersetup{
  colorlinks = true,
            allcolors  = red,
            citecolor  = blue}
\usepackage[compact]{titlesec}
    \titlespacing{\section}{0pt}{1ex}{1ex}
    \titlespacing{\subsection}{0pt}{1ex}{0ex}
    \titlespacing{\subsubsection}{0pt}{0.5ex}{0ex}
\usepackage{soul}
\DeclareUnicodeCharacter{0327}{,}
\DeclareUnicodeCharacter{03BB}{\ensuremath{\lambda}}
\DeclareUnicodeCharacter{039B}{\ensuremath{\Lambda}}

\title{Non-perturbative Thermodynamics of Quark Gluon Plasma and Gravitational Waves}
\author{Narasimha Reddy Gosala \footnote{narasimha.gosala@uleth.ca}, Arundhati Dasgupta \footnote{arundhati.dasgupta@uleth.ca}\\
  Department of Physics and Astronomy, University of Lethbridge,\\ 4401 University Drive, 
  Lethbridge, T1K 3M4,  Canada.}
\date{\vspace{-5ex}}

\begin{document}
\maketitle

\begin{abstract}
Quark–Gluon Plasma (QGP), a strongly interacting state of the early universe, exhibits remarkably fluid-like behavior despite its underlying non-Abelian dynamics. Motivated by these features, we explore time-dependent $SU(2)$ Yang–Mills condensates as non-linear classical background fields to model QGP. We first study quarks in gluon backgrounds and show that quark back-reaction can break the isotropy of the condensate for certain initial conditions. We then compute the one-loop finite-temperature effective action using the background-field method and heat-kernel expansion. The resulting thermodynamic pressure increases with temperature but exhibits an approximately logarithmic dependence. This is expected, as this is the de-confined phase of QGP; it is not exactly an ideal gas due to self-interaction.
We also perform lattice calculations for the system to contrast continuum and lattice perspectives. We then add the GW to the thermodynamic QGP model and show that certain frequencies of the GW can induce instabilities in the QGP.
Our analysis explores the limitations and role of non-perturbative, time-dependent backgrounds in semi-classical description of Yang–Mills dynamics.
 
\end{abstract}

\section{Introduction}

Yang–Mills (YM) fields play a central role in the description of fundamental interactions and provide the theoretical framework for understanding color confinement and chiral symmetry breaking in Quantum Chromodynamics (QCD) \cite{YMint,YMint1,YMint2}. At high temperatures, QCD predicts the formation of a deconfined phase of quarks and gluons known as the quark–gluon plasma (QGP), whose properties are governed by the self-interacting dynamics of YM fields \cite{QGPrev,QGP,QGPbook}. The QGP has been experimentally observed in heavy-ion collisions, offering a direct link between theory and experiment. Prior to thermalization, these collisions are expected to pass through a non-equilibrium phase known as the Glasma, which can be described using classical YM fields \cite{glasma,glasma1}. Beyond high-energy nuclear physics, classical YM fields have also been studied in cosmological contexts, including models of dark energy \cite{DE,DE1,QCDvacuum,QCDvacuum1,QCDvacuum2} and inflationary scenarios such as gauge-flation \cite{gaugeflation,gaugeflation1}.
Our initial motivation to study the interaction of gravitational waves (GW) with classical YM waves \cite{gnr, gnrcon} has thus continued into explorations of YM condensates and QGP in our current paper. In this work, we explore a finite temperature partition function for the QGP and add a GW background to the thermodynamic ensemble.

The QGP, formed both in heavy-ion collisions and in the early universe, has been investigated using a wide range of theoretical approaches, primarily aimed at understanding QCD thermodynamics and the phase transition. Early studies employed the MIT bag model \cite{bagmodel}, followed by effective descriptions such as the NJL and Polyakov–NJL models \cite{NJL,PNJL}. More recently, QGP has also been explored using wave-based approaches \cite{gnr} and condensate models \cite{Prokhorov,con,con2,con3,con4}, in which the plasma is described as a spatially homogeneous Yang–Mills condensate with dynamical fluctuations.

In this context, Prokhorov et al. \cite{Prokhorov} analysed tensor fluctuations around an SU(2) Yang–Mills condensate, and subsequent work extended the analysis to vector fluctuations and to the inclusion of gravitational waves \cite{gnrcon}. Since the QGP contains both gluons and quarks, it is essential to understand fermionic dynamics in such backgrounds. Quark propagation in non-trivial gauge fields has been studied in various settings, including massless fermions \cite{fer}, massive fermions in spontaneously broken gauge theories \cite{fer1,fer2}, and inflationary scenarios \cite{fer3,fer4}. Building on these studies, we extend the analysis of massless quarks in Yang–Mills condensates \cite{gnrcon} by incorporating quark backreaction. We find that the initially isotropic, spatially homogeneous condensate can lose its isotropy in the presence of quarks, highlighting non-trivial quark–gluon interplay in the plasma phase.

Statistical analyses of heavy-ion collisions indicate that strongly interacting matter exists in two distinct phases: a low-temperature confined hadronic phase ($T<T_c$) and a high-temperature deconfined phase known as QGP \cite{QGPrev}. The study of this phase transition has been greatly advanced by lattice QCD simulations at finite temperature \cite{lattice,lattice1}. Various effective and resummed approaches have also been employed to investigate QCD thermodynamics, including the NJL and PNJL models \cite{NJL1,NJL2,NJL3,NJL4,PNJL,PNJL1}, hard-thermal-loop resummation schemes \cite{HTL,HTL1,HTL2}, quasiparticle models \cite{QM,QM1,QM2}, and hadron resonance gas models \cite{HRG}. The bulk thermodynamic properties of the QGP can be characterized by a small set of parameters, most notably the temperature $T$, baryon chemical potential $\mu_B$, and external magnetic fields $B$ \cite{QCDPD}. At high RHIC and LHC energies, the net baryon density is small, and the system is effectively described by a single temperature variable \cite{LHC}. In contrast, experiments at lower energies, such as those at NICA and FAIR, probe regions of the phase diagram with larger chemical potential \cite{LHC1}. While transport properties of the QGP, including shear and bulk viscosities, have also been extensively studied \cite{shear,shearbulk,bulk,bulk1,bulk2,transportphen}, the present work focuses on equilibrium thermodynamic properties.

In this context, QCD phase transitions are commonly characterized by two order parameters: the Polyakov loop for deconfinement and the chiral condensate for chiral symmetry breaking. While models such as NJL and PNJL employ these order parameters to study the transition \cite{NJL1,PNJL}, our analysis instead considers a time-dependent Yang–Mills condensate in the pure deconfined phase, restricting attention to temperatures $T>T_c$ and vanishing chemical potential. The results are therefore most relevant to the deconfined region of the QCD phase diagram, particularly in early-universe applications.

The vacuum of YM theory is an unstable state, and it is believed that the nonlinear nature of YM fields may lead to the formation of condensates which stabilize the vacuum state. There were studies on the stability of different field configurations like the chromomagnetic and chromoelectric fields \cite{magcon,magcon1}. Both of these models suffer from instability, and later many methods were proposed to stabilize these; even the finite temperature extensions do not rectify this \cite{Ebert}. So far, all the studies considering condensates assume the fields to be constant. Here, we consider the condensate to be a spatially homogeneous time-dependent field as in \cite{Prokhorov}. We use the condensate as a background in a finite temperature partition function. We study how the quarks and gauge fluctuations behave in the presence of the background at finite temperature. Modelling the finite temperature action as an analytic continuation of the Lorentzian action, the background action can have a positive or negative sign. The use of the positive action gives an unstable thermodynamic behaviour, which can be used to model dark fluids \cite{dfluid}. These models have been used to derive origins of dark energy and dark matter, and the YM condensate at finite temperature is a candidate model for the same. Adding GW does not change the basic character of the fluid. Using the conventional sign of the action, however, gives stable behaviour of the condensate model of the plasma. We have to assume that the plasma is in equilibrium with a heat bath as in a canonical ensemble. 

 As non-perturbative methods to study YM theory \cite{latqgp} are lattice-based, we extend our work to the lattice. We find an 
 Euclidean lattice action and use the condensate as a background field. The pressure and internal energy behaviour of the SU(2) system is as predicted by earlier papers in lattice gauge theory \cite{su2}. We also find agreement with the calculations of the partition function using heat kernel methods. An interesting behaviour appears when adding GW. The QGP can get unstable due to the addition of GW for certain GW frequencies. This process might induce hadronization and a decay of the condensate as observed in \cite{gnrcon}.
 
 In both the background field method and lattice computations of the thermodynamic quantities, the behavior of the plasma is non-ideal. The YM model is highly interacting, which keeps the pressure above the ideal gas one. This is typical behavior in known observed plasmas where the pressure is due to electrons as well as the ions. Here, the pressure due to the condensates adds to the non-interacting YM pressure. Adding background fluctuations and GW decreases the pressure, signifying that if the plasma decays to its fluctuation modes \cite{gnrcon}, one might obtain a non-interacting ideal gas behavior.
 
 In order to study phase transition in the condensate model, we have to work in a different gauge, as the Polyakov loop is set to 1 in the Hamilton gauge used here. We also expect to study if strangeness is enhanced due to flavor transitions induced by GW, as shown in \cite{gnrcon}, and this is work in progress.

The paper is organized as follows: In section 2 We discuss the dynamics of quarks in the background of an $SU(2)$ Yang-Mills condensate. We discuss a way to build a YM condensate, already introduced in \cite{gnrcon}, in the presence of fermions. Next, in the third section, we quantize the system using the background field method at finite temperatures. Quarks and Gluon fluctuations are added to the partition function, and the heat kernel method is used.  We discuss the Lattice action for the condensate and compute the thermodynamic quantities. In the fourth section, we add GW to the thermodynamic ensemble and find interesting behavior at certain frequencies. Section 5 is a conclusion. We have used Maple for the plots in Secs. 3.5, 3.6 and 4.1, whereas in the remainder of the paper, we used Mathematica.

\section{Quarks in the background of a Yang-Mills condensate}{\label{sec:2}}
Previously in \cite{gnrcon}, we have studied the condensate model of \cite{Prokhorov}, and added quarks to the model. 
In this section, we will start with a review by discussing the solutions of the Dirac equation in the background of the $SU(2)$ YM condensate as in \cite{gnrcon}. Later, we consider the currents built using solutions to the Dirac equation in the background of the condensate and use them as sources of fluctuations over the condensate. 

As given in \cite{gnrcon}, the Dirac equation of a quark interacting with an $SU(2)$ YM condensate is given as 
\begin{equation}{\label{eq:dirac}}
    i \gamma^{\mu} \partial_{\mu} \psi_{\alpha} + \gamma^{\mu} A_{\mu}^a T^a_{\alpha \beta} \psi_{\beta}=0,
\end{equation}
here we absorb the YM coupling constant into gauge field by rescaling: $A^a_\mu \rightarrow \frac{1}{g_{ym}}A^a_\mu$, $T^a$ are the $SU(2)$ generators, $\gamma^{\mu}$ are the four gamma matrices of the Clifford algebra and $\psi_{\alpha}$ are a SU(2) Fermion doublet with $\alpha=1,2$. Considering the SU(2) gauge field as $A^a_0 = 0, A^a_i = \delta^a_i U(t)$, $T^a=\sigma^a/2$, $\sigma^a = \rm{Pauli\; matrices}$ and considering only the left Weyl spinors in $\psi_{\alpha}=(\psi_{\alpha
L} \; \; \; \psi_{\alpha R} )^T$, as $\psi_{1L}=(\psi_{1L1} \; \; \; \psi_{1L2} )^T$, $\psi_{2L}=(\psi_{2L1} \; \; \; \psi_{2L2} )^T$, one gets the following equations:
\begin{align}
    i \partial_0 \psi_{IL1} - \frac{1}{2} U(t) \psi_{IL1}=0, \label{eqn:fermion1}\\
    i \partial_0 \psi_{1L2} -  U(t) \psi_{2L1} + \frac{1}{2} U(t) \psi_{1L2}=0, \label{eqn:fermion2}\\
    i \partial_0 \psi_{2L1} -  U(t) \psi_{1L2} + \frac{1}{2} U(t) \psi_{2L1}=0, \label{eqn:fermion3}\\
    i \partial_0 \psi_{2L2} - \frac{1}{2} U(t) \psi_{2L2}=0. \label{eqn:fermion4}
\end{align}
Taking the condensate as $ U(t)= c_1 {\rm sn}(c_1(t+c_2), -1)$, as in \cite{gnrcon}, the solutions for the above equations are given by
\begin{align}
    \psi_{1L1}(t) &=  A_1 \Lambda^{-1/2},\\
    \psi_{2L2} (t) &= A_2 \Lambda^{-1/2},\\
    \psi_{1L2} (t) &= c_3 \Lambda^{3/2} +c_4 \Lambda^{-1/2},\\
    \psi_{2L1}(t) &= -c_3 \Lambda^{3/2} +c_4 \Lambda^{-1/2},
\end{align}
where $ \Lambda = \left({\rm dn}(c_1(t+c_2), -1)-i \ {\rm cn}( c_1(t+c_2), -1)\right)$, $A_1$, $A_2$, $c_3$ and $c_4$ are integration constants, and $\rm{dn}$ and $\rm{cn}$ are the Jacobi elliptic functions. Note that even though the condensate seems to couple two different flavors ($\psi_{1L2},\psi_{2L1}$) components, however, there is no transition in densities as they are constant in time and space.

As two of the components $\psi_{1L1}, \psi_{2L2}$, are decoupled from the system, we set the constants $A_1=A_2=0$ in the first calculation. The current ($(j^\nu)^a = \bar{\psi}_\alpha\ \gamma^\nu\ T^a_{\alpha\beta}\ \psi_\beta$) generated by these solutions, which are order $g_{ym}$ is found to have the following nonzero components:
\begin{align}
(j^0)^3 &=\frac12 \left(|\psi_{1L2}|^2 -|\psi_{2L1}|^2\right),\label{eq:s1}\\
(j^1)^1 & = - \frac12 \left(\psi^*_{1L2} \psi_{2L1} + \psi^{*}_{2L1}\psi_{1L2}\right),\label{eq:s2}\\
(j^1)^2 & = \frac{i}{2}\left(\psi^*_{1L2} \psi_{2L1}-\psi^*_{2L1}\psi_{1L2}\right),\label{eq:s3}\\
(j^2)^1 & = -\frac{i}{2} \left(\psi^*_{1L2} \psi_{2L1}- \psi_{2L1}^*\psi_{1L2}\right),\label{eq:s4}\\
(j^2)^2 &= - \frac12 \left(\psi^*_{1L2} \psi_{2L1} + \psi^{*}_{2L1}\psi_{1L2}\right),\label{eq:s5}\\
(j^3)^3 &= \frac12 \left(|\psi_{1L2}|^2 + |\psi_{2L1}|^2\right)\label{eq:s6}
\end{align}

These nonzero current densities will act as a source for the Yang-Mills field. Next, we see how the Yang-Mills condensate reacts to the Fermions. This work is to test how the quarks and the gluons `interact' to generate the QGP dynamics.
\subsection{Backreaction of quarks on Yang-Mills fields}
The current densities which we found in the previous section act as a source for the generation of Yang-Mills fields. To understand the backreaction of quarks on the condensate, we will study the Yang-Mills equation of motion with a nonzero source term given as
\begin{equation}{\label{eq:YMeq}}
    \nabla_\mu F^{a \mu\nu} + \  \epsilon^{abc}\  A_{\mu}^{b}\  F^{c\mu\nu} = - g_{ym} (j^{\nu})^a,
\end{equation}
where $F^a_{\mu\nu} = \partial_\mu A^a_\nu - \partial_\nu A^a_\mu + \epsilon^{abc} A^b_\mu A^c_\nu$ is the anti-symmetric field strength tensor of gauge field $A^a_\mu $, $\epsilon^{abc}$ are the structure constants of the gauge group (In the case of SU(2), they are Levi-Civita tensor) and $(j^{\nu})^a$ is Fermion current density which are given in Eqs. (\ref{eq:s1}) - (\ref{eq:s6}).

Taking Hamilton's gauge ($A^a_0=0$) and in the approximation that the gauge fields do not depend on spatial coordinates, one gets the following equations:
\begin{align}
 \epsilon^{abc} A^b_j \partial_0 A^{cj} & = - g_{ym}  (j^0)^a, \label{eq:YMeq0}\\
-\partial_0^2 A^{a}_i + (A^a_j A^b_j A^b_i - (A^b_j A^b_j) A^a_i) & = - g_{ym}  (j_i)^a,\label{eq:YMeqi}\;\; i=1,2,3.
\end{align}
Considering the initial unperturbed solution without Fermions as $A_i^a= U(t) \delta^a_i$ and taking the current induced solutions as perturbations over the condensate, one can write the gauge field ansatz as 
\begin{equation}
A_i^a = U(t) \delta^a_i + \tilde{A}^a_i(t),
\end{equation}
and we solve the back reactions only to the first order in the perturbations. 

If we see the zeroth component of the current ($j^0$), then only its 3rd internal component is non-zero ($(j^0)^3$). Using that, one can rewrite Eq. (\ref{eq:YMeq0}) as
\begin{equation}
    (\partial_0 \tilde{A}^2_1 -\partial_0 \tilde{A}^1_2) U + (\tilde{A}^1_2-\tilde{A}^2_1)\partial_0 U = - g_{ym} (j^0)^3.
\end{equation}
Setting $Z(t)=\tilde{A}_1^2-\tilde{A}_2^1$, one gets a first-order differential equation, which can be integrated. If we take the $U(t)= c_1 {\rm sn}(c_1 t,-1)$, the $(j^0)^3=2 \sqrt{2} \ {\rm sn}^2(c_1 t,-1)$, with $c_3=c_4=1$. The solution for $Z(t)$ is
\begin{equation}
Z(t)= \left(-\frac{2\sqrt{2}g_{ym}}{c_1} t + c_5\right) {\rm sn}(c_1 t,-1).
\end{equation}
This shows that the $A_1^2$ and $A_2^1$ components cannot be equal to make the charge density non-zero for the Fermions.
Next, we solve for the other components of the Gauge field with nonzero current densities.
The Equations for the components are as follows:
\begin{align}
 -\partial_0^2 \tilde {A}_1^2 +  U^2 (\tilde A_2^1 - \tilde A_1^2) =- g_{ym} (j^1)^2, \\
    -\partial_0^2 \tilde {A}_2^1 +  U^2 (\tilde A_1^2 - \tilde A_2^1) = -g_{ym} (j^2)^1, \label{eqn:comp2}\\
-\partial_0^2 \tilde {A}_1^1 - 2  U^2 \tilde A_b^b  = -g_{ym} (j^1)^1,\\
-\partial_0^2 \tilde {A}_2^2 - 2  U^2 \tilde A_b^b  = -g_{ym} (j^2)^2,\\
-\partial_0^2 \tilde {A}_3^3 - 2  U^2 \tilde A_b^b  =- g_{ym} (j^3)^3.
\end{align}
The equation (\ref{eqn:comp2}) can be solved when one plugs in the solution for $Z(t)$ and the current, one gets a differential equation
\begin{equation}
        \partial_0^2 \tilde{A}^1_2= 2 \sqrt{2}\ g_{ym} \  {\rm dn}(c_1 t,-1) {\rm cn}(c_1 t,-1) + U^2 \ Z(t).
\end{equation}
The solution to the above equation is
\begin{equation}
        \tilde{A}^1_2= c_6 + c_7 t - \frac12 Z(t).
\end{equation}

The $\tilde{A}^2_1$ can be determined from the above to be $\tilde{A}^2_1= c_6 + c_7 t + \frac12 Z(t)$, $c_6,c_7$ being the integration constants. We then combine the remaining three equations into an equation for the trace of the gauge field matrix $A_b^b$.
\begin{equation}
          \partial_0^2 \tilde{A}_b^b + 6 U^2 \tilde{A}_b^b = g_{ym} (j_b^b).
\end{equation}  

If we compute the Fermion current diagonal component, they are given as constants, $j_1^1= j_2^2=3/\sqrt{2};$ $ j_3^3= 5/\sqrt{2}$. Then, the equation becomes
\begin{equation}
    \partial_0^2 \tilde{A}_b^b + 6\ c_1^2\ {\rm sn}^2(c_1 t,-1) \tilde{A}_b^b = \frac{11}{\sqrt{2}} g_{ym}.
\end{equation}
We were able to solve the equation analytically. Start by changing the variables by making a substitution $x={\rm am}(c_1 t,-1)$, the resultant equation is solvable. After finding the solution in $x$, we change the variables again to $t$, and the solution turns out to be  
\begin{multline}
    \tilde{A}^a_a (t) = \frac{{\rm cn}(c_1 t,-1){\rm sn}^{1/2}(c_1 t,-1)(2-{\rm cn}^2(c_1 t,-1))^{3/4}}{(2-3\ {\rm cn}^2(c_1 t,-1)+{\rm cn}^4(c_1 t,-1))^{1/4}} \left[ k_1 +\frac{k_2}{4} \left( -\frac{2}{{\rm cn}(c_1 t,-1)}\sqrt{\frac{{\rm cn}^2(c_1 t,-1)-1}{{\rm cn}^2(c_1 t,-1)-2}} \right. \right. \\
    \left. \left. +\sqrt{2}\ F (\sin^{-1} {\rm cn}(c_1 t,-1),\frac12)  \right)  \right] + \frac{11}{2\sqrt{2}} \frac{g_{ym}}{c_1^2} {\rm sn}^2(c_1 t,-1),
\end{multline}
where $k_1,k_2$ are integration constants, $F(x,m)$ is the incomplete elliptic integral of first kind. Since the trace expression has very complicated functions of Jacobi elliptic functions, we could not solve for the diagonal components analytically. However, we found the numerical solutions for $A_1^1,\ A_2^2,\ A_3^3$. Since $j_1^1= j_2^2$, we get the same functional behaviour for $\tilde{A}^1_1$ and $\tilde{A}^2_2$. The solutions are plotted in the following Figs. (\ref{fig:trace},\ref{fig:trace1}).

If we set the initial conditions for the trace function to be zero, then the trace function behaves like the square of the Jacobi function, just as given in the analytical solution (Fig. \ref{fig:trace}). If we take different initial conditions, we see that the trace function keeps on increasing while oscillating. Also, the diagonal components $\tilde{A}^1_1= \tilde{A}^2_2$ show a complementary behavior to $\tilde{A}^3_3$. We observe that the three functions show an oscillatory behavior (Fig. \ref{fig:trace1}). Clearly, with the introduction of the Fermions, the isotropy of the solutions is completely broken. One therefore gets an insight into what the fields might be for QGP.

\begin{figure}[htbp]
    \centering
    \includegraphics[scale=0.8]{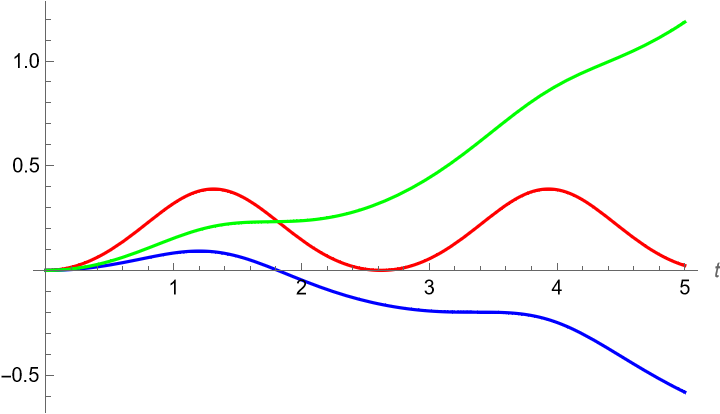}
    \caption{Figure showing the numerical solutions for diagonal components of the gauge field for a short amount of time. The Red, Blue and Green plots corresponds to $\tilde{A}^a_a$, $\tilde{A}^1_1 =\tilde{A}^2_2$ and $\tilde{A}^3_3$, respectively. We choose $g_{ym}=0.1$, $c_1=1$ with initial conditions $\tilde{A}^a_a(0) = \partial_t \tilde{A}^a_a(0) =0$. }
    \label{fig:trace}
\end{figure}

\begin{figure}[htbp]
    \centering
    \includegraphics[scale=0.8]{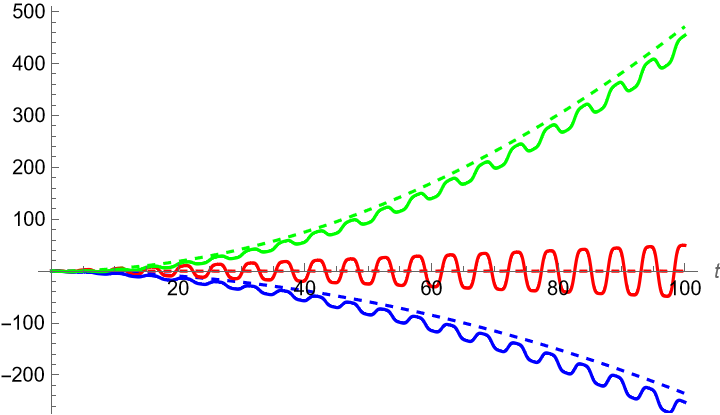}
    \caption{Figure showing the numerical solutions for diagonal components of the gauge field. The Red, Blue and Green plots corresponds to $\tilde{A}^a_a$, $\tilde{A}^1_1=\tilde{A}^2_2$ and $\tilde{A}^3_3$, respectively.The dashed plots corresponds to the initial conditions $\tilde{A}^a_a(0) = \partial_t \tilde{A}^a_a(0) =0$ and the thick lines corresponds to $\tilde{A}^a_a(0) =0$, $ \partial_t \tilde{A}^a_a(0) =1$. We choose $g_{ym}=0.1$ and $c_1=1$. }
    \label{fig:trace1}
\end{figure}

The other components with zero current densities are $\tilde{A}^1_3,\tilde{A}^2_3,\tilde{A}^3_1,\tilde{A}^3_2$. Since these components do not depend on current densities or any of the above components, we can take them to be zero without any inconsistency with the above analysis. As we can see, the components in the $a=1,2$ sector behave differently from the $a=3$ components. We can also keep the $a=1,2$ sector of the $SU(2)$ group equal, but keep the $i,a=3$ sector separate for the system. Thus, our ansatz for the gluon field in the QGP would be
\begin{equation}
A_i^a= U(t)\delta_i^a + \tilde{A}_i^a,
\end{equation}
where $i,a=1,2$; and
\begin{equation}
A_3^3= V(t).
\end{equation}
From this, we can see that the isotropy of the condensate gets broken due to the backreaction of fermions on the condensate. The interest in the above calculation was to investigate the enhancement of strangeness using quark-gluon dynamics. We did not find any such transitions, as it was found using GW in \cite{gnrcon}. Preliminary conclusion can be that the QGP self-interactions require a GW/external interaction to generate strange domination.  In the next section, we extend the idea of using the condensate as classical background fields to evaluate the thermodynamic properties, such as pressure and energy, in the presence of a thermal bath.

\section{Thermodynamic potential in the background of Condensate in an SU(2) model of QCD}
We shall consider an $SU(2)$ gauge model of chiral QCD with the following Euclidean action \cite{Ebert}
\begin{equation}
    S_E = \int d^4x \mathcal{L}_E = \frac{1}{4g_0^2} \int d^4x \ F^{a\mu\nu}\ F^a_{\mu\nu} +\int d^4x \sum_{i=1}^{N_f} \bar{\psi}_i (\gamma^\mu (\partial_\mu + A_\mu )) \psi_i.
    \label{eqn:action}
\end{equation}
We will use the background field method \cite{AbbottBFM,DeWittBFM} to evaluate the 1-loop effective action. In this method, one splits the gauge field ($A^a_\mu$) into a background field ($A^a_\mu$) and quantum fluctuations of the gluon field ($a^a_\mu$)-i.e.,$A^a_\mu \rightarrow A^a_\mu + a^a_\mu $ . In this paper, we take the condensate discussed in the previous section as the background field. As the thermodynamics is obtained using analytic continuation from Lorentzian space, the condensate solution discussed above and as in \cite{gnrcon}. We find that the condensate is an $SU(2)$ instanton in Euclidean space. This is what we discuss next, as well as the sign of the Euclidean action, crucial to the computation of the thermodynamic quantities.

\subsection{Condensate as an Instanton}

In non-Abelian gauge theories, due to the non-linear interaction terms, one gets non-trivial non-perturbative classical solutions of the equation of motion. In Minkowski signature spacetime, these solutions are identified as solitons. In Euclidean space, the solutions are identified as instantons. These Euclidean solutions are also self-dual or anti-self-dual solutions. We define the dual of the gauge field strength in four dimensions using the constant anti-symmetric tensor as
\begin{equation}
\tilde{F}^{\mu \nu}= \frac12 \epsilon^{\mu \nu \rho \sigma} F_{\rho \sigma}.
\end{equation}
In Euclidean signature, the metric, the duality restriction can be imposed as 
\begin{equation}
    \tilde{F}_{\mu \nu}=\pm F_{\mu \nu}.
\end{equation}
It can be shown that the solutions are usually self-dual if 
\begin{equation}
    F_{\mu \nu} \sim \sigma_{\mu \nu}.
    \label{eqn:instan}
    \end{equation}
    as the above commutator of the Pauli matrices is self-dual. 
In the Euclidean solution of the condensate, we find that
\begin{equation}
    U(i\tau)= i c_1{\rm sn} (c_1\tau, -1),
\end{equation}
where $\tau$ is Euclidean time. This is because for the special case of $m=-1$ where $m$ is the parameter of the Elliptic functions, these functions are the Lemniscate Elliptic function. This also means that the condensate is mapped to a purely imaginary value in the Euclidean space.
If one computes the field strength of the above gauge field, one gets
\begin{equation}
    F_{0i}= - \dot{U}(\tau) \epsilon_{iab}\sigma^{ab}; \ \ \ F_{ij}= - U^2 (\tau) \sigma_{ij}.
\end{equation}
This shows that the condensate is actually an $SU(2)$ instanton in Euclidean space, as it can be written in the form of Eq. (\ref{eqn:instan}). 

We use the fact that the Jacobi Elliptic function is periodic in Euclidean time with period $4 K(m)$, to interpret the Euclidean instanton as in equilibrium at temperature $T=c_1/4 K(m)$. Thus, the temperature of the surrounding environment can vary as different instanton solutions will have different constants $c_1$, and the external temperature can be used as a boundary condition to fix the constant $c_1$. 
Note that despite the imaginary nature of the solution, the Euclidean energy of the system is positive definite. If one computes the $\theta$ term using the formula 
\begin{equation}
S_{\theta}= \frac{\theta}{16 \pi^2} \int d\tau \ d^3 x \ \tilde F^{\mu \nu} F_{\mu \nu}= \frac{\theta}{16 \pi^2} V \int_0^{\beta} d \tau \frac{d}{d\tau } U^3(c_1 \tau).
\end{equation}
$V$ is the volume of three space. If $c_1 \tau$ is such that it has the same period as that of the Jacobi function, then the integral is zero, and according to the definition, this would be an instanton with zero winding number. One can have non-trivial winding around the $S^1$, and this is facilitated if $c_1 \tau$ has a fractional period of the Jacobi functions. Thus, in case one is not using finite temperature equilibrium restrictions, the integral will give a non-zero value, and thus the instanton will have a non-trivial winding number.

In the next section, we try to obtain equilibrium thermodynamics using the condensate in a Euclidean background field.

\subsection{The one-loop thermal partition function}
In the background gauge, one gets the following generating functional in Euclidean spacetime
\begin{equation}
    Z[A,j,\eta,\bar{\eta}] = \int \mathcal{D} a \mathcal{D} \psi \mathcal{D} \bar{\psi} \mathcal{D} c \mathcal{D} \bar{c}\; e^{-\int d^4x (\mathcal{L} + j^a_\mu a^a_\mu + \bar{\psi} \eta + \psi \bar{\eta}) },
\end{equation}
where the QCD Lagrangian is given by 
\begin{equation}
    \mathcal{L} = \frac{1}{4g_0^2} (F^a_{\mu\nu}+\hat{\nabla}_\mu a^a_\nu -\hat{\nabla}_\nu a^a_\mu +f^{abc} a^b_\mu a^c_\nu)^2 +\frac{1}{2\xi g_0^2} (\hat{\nabla}_\mu a^a_\mu)^2 - \bar{c} \hat{\nabla}^2 c +  \sum_{i=1}^{N_f} \bar{\psi}_i (\gamma^\mu (\nabla_\mu)) \psi_i.
\end{equation}
Here, $\hat{}\;$ represents terms in adjoint representation and terms without $\hat{}\;$ represents terms in fundamental representation, $\nabla_\mu = \partial_\mu + i A_\mu =\partial_\mu + \frac{i}{2} A^a_\mu \sigma^a$,  $\hat{\nabla}_\mu \equiv \nabla^{ab}_\mu =\hat{\partial}^\mu + \hat{A}_\mu = \delta^{ab} \partial_\mu + f^{acb} A^c_\mu $ is the covariant derivative in the background field, $c$ and $\Bar{c}$ are the ghost fields, and $\hat{\nabla}^2 \equiv (\nabla^2)^{ab} \equiv \nabla^{ac}_\mu \nabla^{cb}_\mu$. We will use Feynman gauge $\xi=1$ for the rest of the calculations. 

For the one-loop calculations, one has to consider the contributions of fluctuations to quadratic order. The gluon Lagrangian is expanded up to quadratic terms in fluctuations as 
\begin{equation}
    \mathcal{L}_g^{(2)} =-\frac{1}{2 g_0^2} a^a_\mu \left(\delta_{\mu\nu} (\nabla^2)^{ab} + 2 f^{acb} F^c_{\mu\nu} \right) a^b_\nu = -\frac{1}{2 g_0^2} a_\mu \left( \delta_{\mu\nu} \hat{\nabla}^2 + 2 \hat{F}_{\mu\nu} \right)a_\nu,
\end{equation}
where we write in the shorthand notation by suppressing color indices, and the corresponding ghost field Lagrangian is 
\begin{equation}
    \mathcal{L}_{gh}^{(2)} = - \bar{c}^a (\nabla^2)^{ab} c^b = -  \bar{c} \hat{\nabla}^2 c,
\end{equation}
and the quark Lagrangian in the one-loop approximation is given by 
\begin{equation}
     \mathcal{L}_q^{(2)} = \sum_{i=1}^{N_f} \bar{\psi}_i (\gamma^\mu  \nabla_\mu)  \psi_i,  
\end{equation}
After doing the Gaussian integrations, one gets the generating functional as \cite{Peskin} 
\begin{equation}
    Z[A] = e^{-\frac{1}{4 g_0^2}\int d^4x ~F^a_{\mu\nu} F^a_{\mu\nu}} \left[{\rm Det}(-\hat{\nabla}^2 \delta_{\mu\nu} - 2 \hat{F}_{\mu\nu})\right]^{-1/2} \left[{\rm Det} (-\hat{\nabla}^2) \right] \prod_{i=1}^{N_f} {\rm Det}\left[ \gamma^\mu \nabla_\mu \right].
\end{equation}
Using the notation $Z=e^{-\Gamma_E}$, the one loop Euclidean effective action is given by 
\begin{equation}
    \Gamma_E = \frac{1}{4g_0^2} \int d^4x ~F^a_{\mu\nu} F^a_{\mu\nu} + \Gamma_g +\Gamma_q = \frac{1}{4g_0^2} \int d^4x ~F^a_{\mu\nu} F^a_{\mu\nu} + \int d^4x ~\mathcal{L}_g + \int d^4x ~\mathcal{L}_q ,
\end{equation}
where 
\begin{align} {\label{eq:gluoneff}}
    \Gamma_g &=\frac{1}{2} {\rm Tr} \ln \left(-\hat{\nabla}^2 \delta_{\mu\nu} - 2 \hat{F}_{\mu\nu} \right) - {\rm Tr} \ln (-\hat{\nabla}^2),  \\ {\label{eq:quarkeff}}
    \Gamma_q &= -N_f {\rm Tr} \ln (\gamma^\mu \nabla_\mu).
\end{align}

For the computation of ${\rm Tr} \ln {\rm K}$ (where K is any operator), we use the Heat Kernel expansion method at finite temperature. In the Heat Kernel method, the trace of an operator is calculated from the coincidence limit of the operator matrix \cite{HK,HK1,HK2}. Consider an operator of form $K = - \nabla^2 + V $, then the trace of $\ln \rm{K}$ is given by 
\begin{equation}
    {\rm Tr} \ln {\rm K} = - \int_0^\infty \frac{d\tau}{\tau} \frac{1}{(4\pi \tau)^{d/2}} H (x,x;\tau),
\end{equation}
where $H(x,x;\tau) = \int d^dx~\sum_{n=0}^\infty \tau^n {\rm tr}(b_n) $ is the Heat kernel corresponding to the operator K and $b_n$ are the Seeley-deWitt coefficients at zero temperature. Here, $d$ is the number of dimensions, and $\rm {tr:=tr_c tr_d tr_l}$ denotes the traces over color, Dirac, and Lorentz indices. To evaluate thermodynamic potential, one has to find the Seeley-deWitt coefficients at finite temperatures. At finite temperatures, the expression for $H(x,x:\tau)$ gets an overall factor term \cite{HKt}. Later, it is found that the expression is incomplete, and this incompleteness is due to the $A_0$ term behaving differently at finite temperature \cite{HKt2}. The corrected heat kernel expansion of the ${\rm Tr} \ln \rm{K}$ is given as \cite{HKt1} 
\begin{equation}
    {\rm Tr} \ln {\rm K} = - \int_0^\infty \frac{d\tau}{\tau} \frac{1}{(4\pi \tau)^{d/2}} \int d^dx \sum_{n=0}^\infty \tau^n b_n^T(x), 
\end{equation}
with the following Seeley-deWitt coefficients at finite temperature (up to mass dimension 4) 
\begin{align}
    b_0^T(x) & = {\rm tr} (\varphi_0 b_0) = {\rm tr} (\varphi_0)\\
    b_1^T(x) & = {\rm tr} (\varphi_0 b_1) = {\rm tr} (-\varphi_0 V)\\
    b_2^T(x) & = {\rm tr} \left(\varphi_0 b_2 -\frac{1}{6} \Bar{\varphi}_2 E_i^2\right) = {\rm tr} \left(\frac{1}{2} \varphi_0 V^2- \frac{1}{3} \varphi_0 E_i^2 + \frac{1}{12} \varphi_0 F^2_{ij}\right),
\end{align}
where $b_0,b_1,b_2$ are Seeley-deWitt coefficients at zero temperature,$\bar{\varphi}_2 = \varphi_0+2\varphi_2 $, $E_i = F_{0i}$ and 
\begin{align}{\label{eq:ploop}}
    \varphi_n^\pm(L) &= (4\pi\tau)^{1/2} \frac{1}{\beta} \sum_{p_0^\pm} \tau^{n/2} R^n e^{\tau R^2}, \; \; R=i p_0^\pm -\frac{1}{\beta} \ln(L),\; \beta=\frac{1}{T},\\ 
    p_0^+ &= \frac{2\pi k}{\beta} \; ({\rm for\; bosons}), \; p_0^- = \frac{2\pi}{\beta}\left( k+\frac12\right) \; ({\rm for\; fermions}),
\end{align}
with $L$ being the untraced Polyakov loop given by 
\begin{equation}
    L = T \exp \left[-\int_{x_0}^{x_0+\beta} A_0(x_0',x) dx_0' \right].
\end{equation}

Before we move on to the actual calculation of functional determinants, we consider the background gauge field to be a spatially homogeneous and isotropic condensate field with the following ansatz
\begin{equation}
    A_0^a = 0,\; \; A^a_i = \delta^a_i U(t)
\end{equation}
where $U(t)$ is the solution of the classical Yang-Mills equations of motion given in Sec. \ref{sec:2}.

With the above ansatz, one can see that the $A_0 = 0 \implies L = I \;({\rm Identity\; in\; color\; space})$. Derivations for effective Lagrangian contributions from quarks (\ref{eq:quarkeff}) and gluons (\ref{eq:gluoneff}) are given in Appendix \ref{app:1}. Here, we present the results for $\mathcal{L}_g$ and $\mathcal{L}_q$, upto mass dimension 4, 
\begin{multline}
    \mathcal{L}_g  = - \frac{\pi^2}{45} T^4 (N^2-1) - \frac{1}{(4\pi)^2} \frac{11}{12} \left[\frac1\epsilon +\gamma_E + \ln(4\pi) \right] {\rm tr_c} (\hat{F}^2_{\mu\nu}) -\frac{11}{6} \frac{1}{(4\pi)^2} \ln\left( \frac{\Lambda}{4\pi T}\right) {\rm tr_c} (\hat{F}^2_{\mu\nu})\\
     +\frac{1}{3} \frac{1}{(4\pi)^2} {\rm tr_c}(\hat{E}^2_i).
\end{multline}
In the above $\epsilon\rightarrow 0$ is a small parameter introduced for dimensional regularization of the partition function, $\Lambda$ is the energy scale in the renormalization, and $N_f$ is the number of fermions in the system. 
\begin{multline}
    \mathcal{L}_q = -\frac{7\pi^2}{180} N N_f T^4 + \frac{1}{(4\pi)^2} \frac{N_f}{3} \left[\frac1\epsilon +\gamma_E + \ln(4\pi) \right] {\rm tr_c} (F^2_{\mu\nu}) +\frac{N_f}{3} \frac{1}{(4\pi)^2} \left[2 \ln\left( \frac{\Lambda}{4\pi T}\right) + 2 \ln4 \right] {\rm tr_c} (F^2_{\mu\nu})\\
    - \frac{1}{(4\pi)^2} \frac{2}{3} N_f  {\rm tr_c}(E^2_i).
\end{multline} 

As we can see, there are some divergent terms in those expressions. After treating the divergences by absorbing them in the bare coupling constant (See Appendix \ref{app:1}), we get the final renormalised Lagrangian as 
\begin{multline}
    \mathcal{L} = - \frac{\pi^2}{45} T^4 (N^2-1)-\frac{7\pi^2}{180} N N_f T^4 +\left[\frac{1}{4g^2(\Lambda)} - \frac12 \beta_1 \ln\left(\frac{\Lambda}{4\pi T}\right) + \frac{1}{(4\pi)^2} \frac{N_f}{3} \ln 4 \right] F^{a\mu\nu} F^a_{\mu\nu} \\
     - \frac13 \frac{1}{(4\pi)^2} (N_f-N) E^{ai} E^a_i
\end{multline}

where $\beta_1= ((11/3)N-(2/3)N_f)/4\pi^2$, the details of which have been derived in the Appendix \ref{app:1}. Next, we derive the thermodynamic potential for the QGP.
\subsection{Thermodynamic Potential and other Thermodynamic functions}
With a known partition function, we obtain the thermodynamic potential as a function of temperature as  
\begin{equation}
    \Omega(T) = -\frac{T}{V} \ln Z(T).
\end{equation}
Using the Lagrangian we found, we write the thermodynamic potential as 
\begin{multline}{\label{eq:thdpot}}
    \Omega(T) = - \frac{\pi^2}{45} T^4 (N^2-1)-\frac{7\pi^2}{180} N N_f T^4 + \frac{T}{V} \left[\frac{1}{4g^2(\Lambda)} -\frac12 \beta_1 \ln\left(\frac{\Lambda}{4\pi T}\right) + \frac{1}{(4\pi)^2} \frac{N_f}{3} \ln 4 \right] \left[ \int d^4x ~F^{a\mu\nu} F^a_{\mu\nu}\right]\\
     - \frac{T}{3V} \frac{1}{(4\pi)^2} (N_f-N) \left[\int d^4x ~E^{ai} E^a_i\right].
\end{multline}
The measurable quantities of the system like pressure $P$, entropy density $S$, energy density $\epsilon$, trace anomaly $\Theta^\mu_\mu$ and speed of sound $c^2_s$ are related to $\Omega$ by 
\begin{align}
     P & =-\Omega,\; \; \; S=-\frac{\partial \Omega}{\partial T},\; \; \; \epsilon = -P + T S,\\
     \Theta^\mu_\mu & = \epsilon - 3 P,\;\;\; c^2_s \equiv \frac{\partial P}{\partial \epsilon}.  
\end{align}
As we discussed before, we considered the background gauge field to be a spatially homogeneous and isotropic condensate ($A^a_0=0, A^a_i= \delta^a_i U(t)$). With this background, we found the thermodynamic potential to be 
\begin{multline}
    \Omega (T) = - \frac{\pi^2}{45} T^4 (N^2-1)-\frac{7\pi^2}{180} N N_f T^4 - 2 \left[\frac{1}{4g^2(\Lambda)} - \frac12 \beta_1 \ln\left(\frac{\Lambda}{4\pi T}\right) + \frac{1}{(4\pi)^2} \frac{N_f}{3} \ln 4 \right] \\
    c_1^3 \left( c_1 - 2 i T {\rm cn}(c_1c_2,-1){\rm sn}(c_1c_2,-1){\rm dn}(c_1c_2,-1)+ 2i T {\rm cn}(c_1(c_2-\frac{i}{T}),-1){\rm sn}(c_1(c_2-\frac{i}{T}),-1) \right.\\
    \left. {\rm dn}(c_1(c_2-\frac{i}{T}),-1) \right) + \frac13 \frac{1}{(4\pi)^2} (N_f-N) c_1^3 \left( 2 c_1 -  i T {\rm cn}(c_1c_2,-1){\rm sn}(c_1c_2,-1){\rm dn}(c_1c_2,-1) \right. \\
    \left. + i T {\rm cn}(c_1(c_2-\frac{i}{T}),-1){\rm sn}(c_1(c_2-\frac{i}{T}),-1){\rm dn}(c_1(c_2-\frac{i}{T}),-1) \right),
\end{multline}
where $c_1$ and $c_2$ are two arbitrary constants fixed by the initial conditions of the elliptic Jacobi differential equation. Similarly, one can find the thermodynamic quantities using $\Omega$.

Before discussing the results, we consider the one-loop running coupling constant. By re-writing the coupling constant at some energy scale as \cite{Collins,Kapusta} 
\begin{equation}
    \alpha_s(\Lambda) \equiv \frac{g^2(\Lambda)}{4\pi} = \frac{1}{4\pi \beta_1 \ln(\Lambda^2/\Lambda^2_{\bar{MS}})}.
\end{equation}
One can fix the critical scale $\Lambda^2_{\bar{MS}}$ which determines the phase transition, by knowing the running coupling constant value at some energy scale $\Lambda$.  In particular, by setting $\alpha_s(1.5 {\rm GeV}) = 0.326$ \cite{latticeg}, we can fix the $\Lambda_{\bar{MS}} = 176\ {\rm MeV}$ for one-loop calculations. Since we are considering $SU(2)$, we cannot directly use this value, so we consider a certain range for $\Lambda_{\bar{MS}}$. We choose the renormalization scales to be $\Lambda=2\pi T$. 

\begin{figure}[htbp]
\centering
    \begin{subfigure}{0.4\textwidth}
    \centering
      \includegraphics[width=\textwidth]{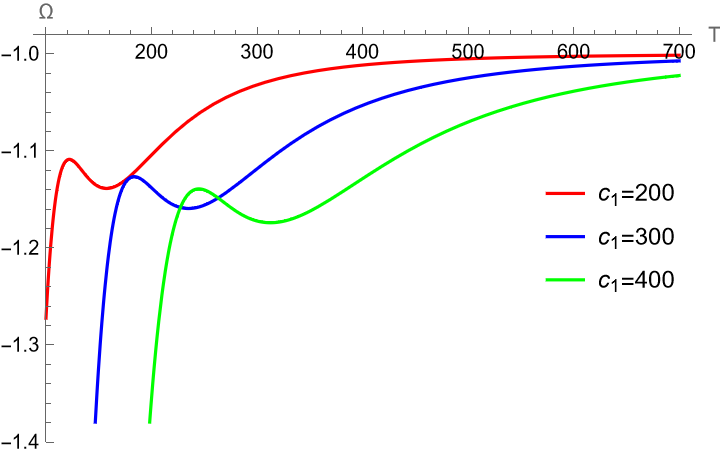}
      \caption{$\Omega$ as a function of $T$}
      \label{fig:thdpotdiffc1}
    \end{subfigure}
    \hfill
    \begin{subfigure}{0.4\textwidth}
    \centering
      \includegraphics[width=\textwidth]{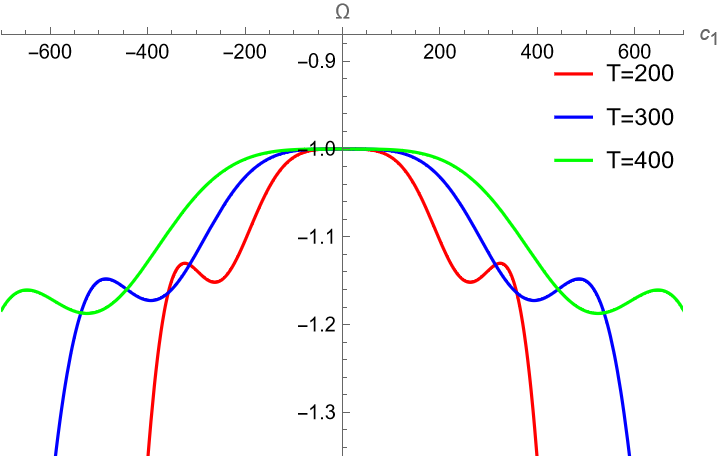}
      \caption{$\Omega$ as a function of $c_1$}
      \label{fig:thdpotdiffT}
    \end{subfigure}
    \hfill
    \caption{Figures showing the thermodynamic potential $\Omega$ as a function of $c_1$ and temperature $T$. For this, we set $\Lambda_{\bar{MS}} = 200 $ MeV. }
    \label{fig:thdpotcon1}
\end{figure}

If one considers the condensate as a separate non-dynamical background field away from the thermodynamic equilibrium, we can treat the constants $c_1$ and $c_2$ as individual parameters. Then, the thermodynamic potential is a function of $c_1$ and $c_2$ along with $T$. For simplicity, we first consider the effect of only the dependence of $\Omega$ on $c_1$. The thermodynamic potential shows an inflationary type behavior as a function of temperature for each particular $c_1$ value as seen in (Fig. \ref{fig:thdpotdiffc1}). With increasing $c_1$ value, the point of local minimum or false vacuum moves toward the origin. As one can see, the system first reaches a false bottom/vacuum as it moves toward minima and then, finally, reaches a true vacuum in which the thermodynamic potential is minimum. This behavior is the same as that of the  inflationary theory. The main difference is that one gets the inflationary type potential without introducing any scalar particle, and with gauge fields alone. This line of study, using gauge fields to drive inflation, has been researched by many authors \cite{gaugeflation}. As we discussed earlier, this scenario happens when one treats the condensate as an background field independent of the ambient temperature.

\begin{figure}[htbp]
    \centering
    \includegraphics[width=0.7\linewidth]{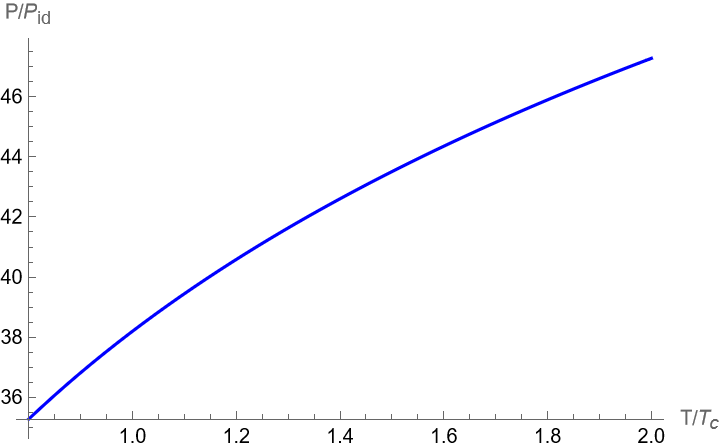}
    \caption[Figures showing the normalized pressure as a function of $T/T_c$]{Figures showing the normalized pressure as a function of temperature $T/T_c$. For this, we set $N=2$, $N_f =2$ and $\Lambda_{\bar{MS}} = 200 $ MeV.}
    \label{fig:presscon}
\end{figure}

 If we treat the condensate along with fluctuations in the thermodynamic equilibrium, then one has to modify the variables $c_1$ and $c_2$ to accommodate equilibrium at a given temperature. This is because, to match the Matsubara frequencies (interpreting the period of fields in imaginary time as inverse temperature), we have to do a one-to-one mapping of the period of Jacobi functions to temperature. After doing that, one finds $c_1 = 4 i K(-1) T $, where $K(m)$ is the Complete Elliptic integral of the first kind, while $c_2$ is still independent of temperature. Now, the thermodynamic potential ($\Omega$) is a function of temperature $T$ and $c_2$. After some analysis, we found that the terms dependent on $c_2$ will cancel each other, leaving the thermodynamic potential as a function of temperature $T$. From this, we calculate the other thermodynamic quantities such as pressure, as shown in Fig. \ref{fig:presscon}. 

\begin{multline}
    \Omega (T) = - \frac{\pi^2}{45} T^4 (N^2-1)-\frac{7\pi^2}{180} N N_f T^4 + 2 \left[-\frac{1}{4g^2(\Lambda)} +\frac12 \beta_1 \ln\left(\frac{\Lambda}{4\pi T}\right)-\frac{1}{(4\pi)^2} \frac{N_f}{3} \ln 4 \right](4iK(-1) T)^4 \\
    + \frac13 \frac{1}{(4\pi)^2} (N_f-N) 2 (4 i K(-1) T)^4 .
\end{multline}

\begin{figure}[htbp]
    \centering
    \includegraphics[width=0.5\linewidth]{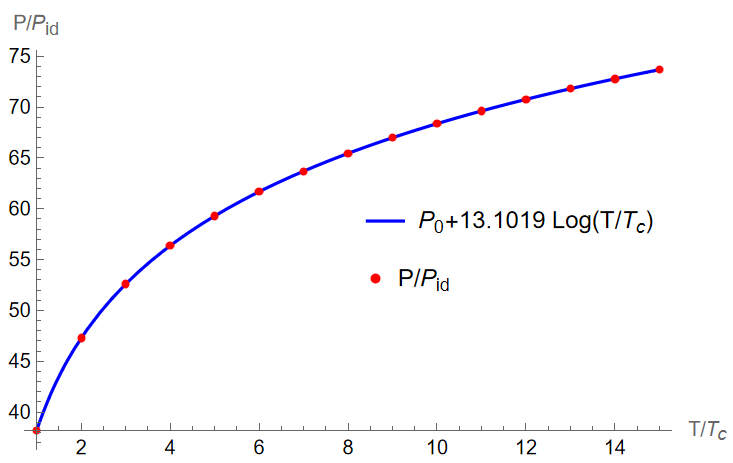}
    \caption{Comparison plot of normalised pressure with ${\rm Log} (T/T_c)$.}
    \label{fig:Presscomplog}
\end{figure}

Figure \ref{fig:presscon} shows the normalized pressure $P/P_{id}$ as a function of $T/T_c$ for the time-dependent condensate background, where $P_{id} $ is the idealized pressure. The pressure increases with temperature, but its temperature dependence (Fig. \ref{fig:Presscomplog}) is approximately logarithmic and can be fitted by
\begin{equation}
    \frac{P}{P_{id}} \approx a + b \ \log\left(\frac{T}{T_c}\right), \  \ P_{id} =  \frac{\pi^2}{45} T^4 (N^2-1) + \frac{7\pi^2}{180} T^4 N N_f,
\end{equation}
with constants $a$ and $b$. This behaviour differs qualitatively from lattice Yang–Mills thermodynamics, where $P/P_{id}$ rises sharply around $T_c$ and then approaches the Stefan–Boltzmann limit ($\propto T^4$) from below at high temperature. In our case, the normalized pressure does not saturate unity but instead continues to grow slowly with $T$.

The origin of this discrepancy is the dominant classical contribution of the background field. The periodicity condition fixes the condensate amplitude to scale as $c_1 = 4 K(-1) T$, causing the field strength to scale as $F^2_{\mu\nu} \propto c_1^4 \propto T^4$ with a large numerical prefactor. Since the thermodynamic potential contains a term proportional to  $F^2_{\mu\nu} / g^2 $, this classical action density becomes much larger than the Stefan–Boltzmann pressure. After normalizing by the ideal-gas value $P_{id} \propto T^4$, the remaining temperature dependence arises primarily from the logarithmic contributions in the one-loop determinants, leading naturally to the observed form $P/P_{id} \propto \log(T/T_c) $.

Physically, this indicates that although the condensate is a valid non-trivial classical solution, it does not correspond to the thermodynamically preferred phase of the finite-temperature de-confined Yang–Mills ensemble in lattice simulations. The equilibrium configuration sampled in lattice simulations has a much lower classical action density, which explains why lattice pressure approaches the ideal-gas limit rather than exhibiting logarithmic growth. Our results therefore, highlight that the present background should be interpreted as a semi-classical probe of Yang–Mills dynamics rather than an ideal gas description of the equilibrium equation of state.

By contrast, the spectral approach used in \cite{nonpert} models the thermal system as an ensemble of massive scalar excitations with non-zero momentum. This construction allows the pressure to be tuned to existing lattice data and yields good phenomenological agreement. However, the two approaches are conceptually different: the spectral model is designed to reproduce the equation of state, whereas our background-field setup is a semi-classical probe of Yang–Mills vacuum structure.

\noindent\textbf{Comparison with existing Lattice Results:}

The background-field method combined with the heat-kernel expansion offers a controlled way to evaluate quantum corrections around a fixed classical configuration, but it does not naturally correspond to the thermodynamic equilibrium phase sampled in lattice simulations. In our case, the periodicity condition forces the condensate amplitude to grow linearly with temperature, leading to a large classical action density and consequently to a normalized pressure that exceeds the Stefan–Boltzmann limit.  In contrast, spectral methods such as \cite{nonpert} assume an infinite tower of thermal excitations and allow parametric matching to lattice data by construction. Thus, the disagreement in magnitude between our results and previous lattice simulations reflects the different physical assumptions underlying the two frameworks, rather than a failure or inconsistency of the background-field approach. In the next section, we will consider the same analysis for the modified condensate ansatz: YM field generated by the back reaction of quarks on the condensate.

\subsection{Thermodynamic Potential with the modified Condensate ansatz}
\begin{figure}[htbp]
    \centering
    \includegraphics[width=0.5\linewidth]{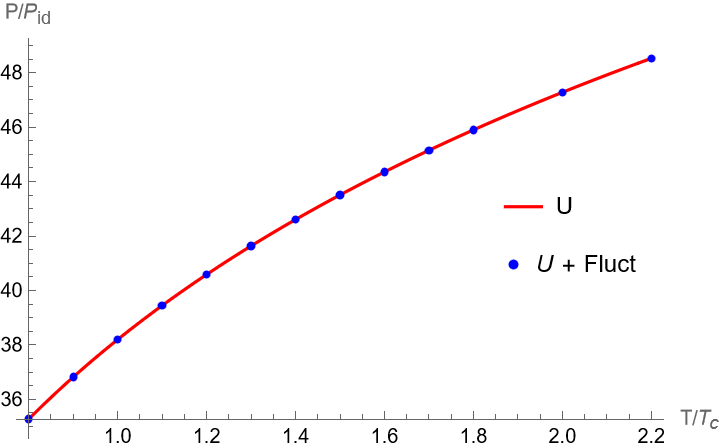}
    \caption{Figure comparing pressure with and without the fluctuations. }
    \label{fig:pressmodcon}
\end{figure}

In this section, we consider the modified condensate ansatz, which is a modification of the original ansatz due to the backreaction of quarks (Sec: \ref{sec:2}):
\begin{equation}
    A^a_0=0:\; \; A^a_i = \delta^a_i U(t) + \tilde{A}^a_i(t).
\end{equation}
The form of thermodynamic potential is the same as in the previous case, Eq. (\ref{eq:thdpot}). Although with the modification of ansatz, the form of $F^{a\mu\nu}F^a_{\mu\nu}$ changes to 
\begin{multline}
    F^{a\mu\nu} F^a_{\mu\nu} = 2 \left[ \left((\tilde{A}^1_2)^2 + (\tilde{A}^2_1)^2 +(\tilde{A}^1_1 + U)^2 +(\tilde{A}^2_2 + U)^2 \right) (\tilde{A}^3_3+U)^2 + \left(\tilde{A}^1_2 \tilde{A}^2_1 - (\tilde{A}^1_1 + U)(\tilde{A}^2_2+ U )\right)^2  \right.\\
   \left.  + (\partial_t \tilde{A}^1_2)^2 + (\partial_t \tilde{A}^2_1)^2 + (\partial_t U + \partial_t \tilde{A}^1_1)^2
    +(\partial_t U + \partial_t \tilde{A}^2_2)^2 +(\partial_t U + \partial_t \tilde{A}^3_3)^2   \right],
\end{multline}
and the form of $E^a_i E^{ai}$ changes to 
\begin{equation}
    E^a_i E^{ai} = (\partial_t \tilde{A}^1_2)^2 + (\partial_t \tilde{A}^2_1)^2 + (\partial_t \tilde{A}^1_1 + \partial_t U)^2 + (\partial_t \tilde{A}^2_2 + \partial_t U)^2 + (\partial_t \tilde{A}^3_3 + \partial_t U)^2.
\end{equation}
With these changes, we found the thermodynamic potential and other thermodynamic quantities. Since we are not able to find the analytical solutions for some of the perturbation functions ($\tilde{A}^1_1, \tilde{A}^2_2, \tilde{A}^3_3 $), we find the numerical solutions for the thermodynamic quantities. The plot for the pressure with the condensate ansatz in the presence of fermions is given in Figs. (\ref{fig:pressmodcon}).

As is evident, the addition of the quarks and use of perturbative back reaction show the same behavior as that predicted using the background condensate for the QGP thermodynamics. We next explore lattice gauge theory, and if the use of Lattice methods introduces new physics, which is non-perturbative in the gauge gluon coupling.

\subsection{A Lattice Contribution}
In this section, we discuss a lattice calculation of the gluon plasma thermodynamic potential. Adding quarks in the lattice introduces additional complications, including fermion doubling \cite{fermlat}, which we avoid at this time. In a lattice, the degrees of freedom of the system comprise the Wilson line and the discrete version of the local gauge action defined on a plaquette. Given a Gauge field $A_{\mu}$, and a lattice with links of length $a$, the Wilson line is  
\begin{equation}
    U_{\mu}(x) = \exp\left(i a A_{\mu}\right),
    \label{eqn:wiline}
\end{equation}
with $x$ specifying the starting location of the link. 
The plaquette comprises four links, enclosing a rectangular area. The plaquette function $W_{\mu \nu}$ is defined as
\begin{equation}
    W_{\mu \nu}= {\rm Tr}\  U_{\mu} (x) U_{\nu} (x+\mu) U^{\dag}_{\mu}(x+\nu) U^{\dag}_{\nu}(x), 
    \label{eqn:plaq}
\end{equation}
The lattice action for one plaquette $ S(W_{\mu \nu})$, is obtained from the plaquette function. The finite temperature lattice partition function is given as

\begin{equation}
    Z= \int {\cal D} U_{\mu} \exp\left(- \sum_{\rm plaq}S(W_{\mu \nu})\right).
\end{equation}

 We take a cubic lattice in 3 spatial dimensions. We take Euclidean time periodic in $\beta$, as in finite temperature field theory. As the condensate is time-dependent, we transform the solution from Lorentzian space-time to Euclidean time; however, the Euclidean time is kept continuous. 
\noindent
As we are computing thermodynamic quantities, and the action is analytically continued from the Lorentzian one, it is important to understand the correct sign of the Euclidean lattice action.  If one computes the action of the Lorentzian metric condensate obtained in \cite{gnrcon}, in the continuum case, one finds it is 
\begin{equation}
    S_{\rm Lor}= -\frac{3 V}{8 g^2} \int (1- 2 U^4) dt 
\end{equation}
where $V$ is the spatial volume. If we Euclideanize the above by changing $t\rightarrow -i \tau$, where $\tau$ is the Euclidean time, 
\begin{equation}
    S_{\rm Eucl}= \frac{3V}{8 g^2} i \int(1- 2 U_I^4) d \tau
    \label{eqn:euclidact}
\end{equation}
As the solution for $U(t)$ is the Lemniscate function, the $t\rightarrow -i\tau$ transformation is implemented as $U(t)\rightarrow i U(\tau)$ and therefore the $U^4$ term remains unchanged in sign in the Wick rotated action. There is an ambiguity in the Wick rotation process, as $t\rightarrow \pm i \tau$ would essentially achieve the same result for the transformation of the Lorentzian metric to the Euclidean metric.  Keeping in mind that, if we take the Euclidean action as in Eq. (\ref{eqn:euclidact}), we choose the sign for the Euclidean lattice action which gives the same sign for the $U^4$ term as in Eq. (\ref{eqn:euclidact}).

We take the plaquettes along the faces of a cube with sides $a$. The sides of the cube are along the $x-y-z$ axes. As the Gauge connection is taken to be time dependent only, the plaquettes along all three $x-y$, $x-z$, $y-z$ planes give the same contribution to the action. If we had taken a 4-dimensional hyper-Lattice, then the contribution from the $t-x$, $t-y$, $t-z$ plaquettes to the Wilson action would be constants due to the Hamilton gauge, in which $A_0^a=0$.
To build the lattice action, let us take the $x-z$ plane plaquette as an example. Using Eqs. (\ref{eqn:wiline} \& \ref{eqn:plaq}), one gets,
\begin{equation}
    W_{xz}= {\rm Tr}\left[ \ \exp\left(i a\  U \frac{\sigma^1}{2}\right)\exp\left(i a\ U \frac{\sigma^3}{2}\right) \exp\left(- i a\ U^* \frac{\sigma^1}{2}\right)\exp\left(-i a\ U^* \frac{\sigma^3}{2}\right)\right] 
\end{equation}
Using the identity
$$\exp\left(i\  \vec{x}\cdot\vec{\sigma}\right)= \cos(x)+ i\frac{\vec{x}\cdot\vec{\sigma}}{x}\sin (x)$$
where $x=\sqrt{x_1^2+x_2^2+x_3^2}$, one can obtain the following

\begin{multline}
    W_{xz}= {\rm Tr} \left[ \left(\cos\left(\frac{aU}{2}\right) + i \sigma^1 \sin\left(\frac{aU}{2}\right)\right)\left(\cos\left(\frac{aU}{2}\right) + i \sigma^3 \sin\left(\frac{aU}{2}\right)\right) \right. \\
     \left. \left(\cos\left(\frac{aU^*}{2}\right)- i \sigma^1 \sin\left(\frac{aU^*}{2}\right)\right)\left(\cos\left(\frac{aU^*}{2}\right)-i \sigma^3 \sin\left(\frac{aU^*}{2}\right)\right) \right]
\end{multline}

Note that the $U$ function is imaginary in Euclidean time. This therefore gives the trigonometric ratios as Hyperbolic functions of this $U_I$ imaginary component. The Wilson line defined in Eq. (\ref{eqn:wiline}) ceases to be unitary in that case, and we have to redefine the plaquette action.

If we use the action for one plaquette as given in \cite{Kapusta}(with the correct sign), we get 
\begin{eqnarray*}
    S(W_{xz})& = & -\frac{4}{g^2} \left(1- \frac12 {\rm Tr} W_{xz}\right)\\
    & =&- \frac{8}{g^2} \left( \sinh^4 \left(\frac{aU_I}{2}\right) + 2\sinh^2\left(\frac{aU_I}{2}\right)\right)
\end{eqnarray*}
    
As there is a contribution only from the time-dependent $U(t)$, all the spatial plaquettes give the same contribution. This gives an overall volume factor in the total action from the sum over the spatial plaquettes. Note that the $U$ function is imaginary in Euclidean time, which has a period of $\beta$. Thus, the total action is actually
\begin{equation}
S_{\rm Lat}= -\frac{24 m}{ g^2} \int_0^{\beta} \left( \sinh^4 \left(\frac{a U_I}{2}\right) + 2 \sinh^2\left(\frac{a U_I}{2}\right)\right) ~ d \tau
\end{equation}

The number of plaquettes, including the boundary ones, is $m$ from each of the 3 sides. We keep time as a continuum with a period of $\beta$. To get an analytic understanding of the system, we expanded the hyperbolic functions to order $a^4$ as
\begin{equation}
S_{\rm Lat}= -\frac{3 m}{g^2} \int_0^{\beta} \left(\frac{5 U_I^4 a^4}{6} + 4 a^2\  U_I^2\right) \ d\tau, 
\end{equation}
Note the interesting appearance of the $U_I^2$ term, which is not there in the classical Euclidean action. The obvious reason is that we have used the same formula as for real Gauge fields. In principle, the formula for the plaquette action should be such that it is adapted for the case that the analytically continued Euclidean gauge field is imaginary. In which case, we postulate two options for the Wilson line operator, one of which is straightforward:
\begin{equation}
U_{\mu}= \exp\left(a A_{\mu} \right).
\end{equation}
In which case the operator is unitary, and $U^{\dag}_{\mu}=U^{-1}_{\mu}$, otherwise, one should use $U^{-1}_{\mu}$  for the reverse links in the plaquette action if the Wilson operator is non-Unitary. If we use that, the $U^2$ term gets cancelled, but the $U^4$ term remains with the same sign as calculated above. 

In either of the options mentioned above to get the correct plaquette action, one gets the $U_I^4$ term, same as the one obtained above.
The action for one plaquette, therefore, has
\begin{equation}
    S(W_{xz}) =-\frac{8}{g^2} \int_0^{\beta} \sinh^4\left(\frac{a}{2} c_1 {\rm sn}(c_1 \tau,-1)
\right) d \tau .
\end{equation}

To facilitate the periodicity of the function, we use $c_1= 4K(-1)/\beta$, which ensures that ${\rm sn}(c_1 (\tau+\beta),-1)={\rm sn}(c_1 \tau, \beta)$.   In the event $c_1a\approx 4 K(-1)\  T\ a\approx 1$, the usual Lattice scale of $10^{-15} {\rm m}$ is for studying confined systems. As we are studying the de-confined phase, the Lattice spacing is taken as smaller than that scale. In fact, if we set $T\sim T_c; a\sim 1/T_c$, 
the integral is exactly calculable, and the action scales as $1/T$. If we, on the other hand, make a small parameter of $a T$, we obtain the Lattice action as: 
\begin{equation}
    S{(W_{xz})}= -\frac{1}{2 g^2} 1.748032 (4 K(-1))^3 T^3 a^4.
\end{equation}
This is the action due to one plaquette, but as the $U$ function is independent of the spatial coordinates, plaquettes in the $W_{yz}, W_{yx}$ would give exactly the same contribution to the Lattice action as above. 

Let $N$ be the total number of cubes of side $a$ which are used to fill the 3-D Lattice and is different from the number of plaquettes. If one takes $n$ as the number of vertices on each $x,\ y,\ z$ dimensions, then the total number of plaquettes, including the boundary one, is:
\begin{equation}
    m=n^3-2 n^2+n; \ \ \ N= (n-1)^3.
\end{equation}

\begin{figure}[htbp]
    \centering
   \includegraphics[width=0.7\linewidth]{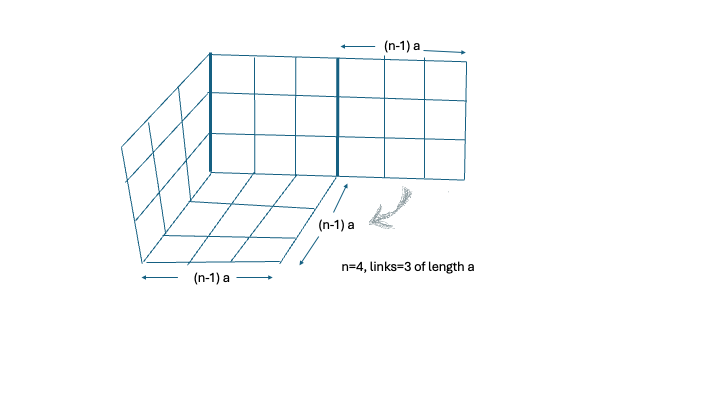} 
   \caption{3-D Lattice and the number of plaquettes}
\label{fig:plaq}
\end{figure}

We calculate this as described using Fig. (\ref{fig:plaq}): From a representative lattice with 4 vertices on each side, we find $3^2$ plaquettes on each face of the boundary cube of sides $3a$. If we then count the number of these \say{faces} which layer the inside of the cube,   perpendicular to one face, one gets the number $3^2\times 4=36$. As the system is symmetric, the same count will apply for the inside faces in the direction perpendicular to the other faces. For a generic $ n$-vertex lattice, the total count is therefore  $3 n(n-1)^2= 3(n^3-2n^2+n)=3 m$. 
Note that the total number of cubes of each volume $a^3$ is $N=(n-1)^3$, which gives a total spatial volume: $V = (n-1)^3 a^3$.

If we calculate the Free energy, Pressure and energy density of the above system, it will scale with $T^4$, commensurate with what is expected from a de-confined gas. To obtain that, we observe that the partition function is given as
\begin{equation}
    Z= \exp(-S_{\rm Lat}).
\end{equation}
As this is a pure fixed background, there are no sums or integrals involved. When one adds the fluctuations over the background, one will have to integrate over these to get the partition function.
When we find the free energy, we use the formula:
\begin{equation}
    \Omega= - T \ln Z = - \frac{0.874017}{g^2} 3 m (4 K(-1))^3 T^4 a^4 .
\end{equation}

The above is almost the same as that of a gas of free particles, and one can envisage a `deconfined phase' of quarks and gluons.
However, notice that we are discussing the partition function above a certain energy scale, and therefore, one should add to the above the effect of renormalisation of the coupling constant up to this energy scale. We add the flow of the coupling constant to the partition function.  Thus:  
\begin{equation}
\frac1{g^2(\Lambda)}= \beta_1 \ln \left(\frac{\Lambda^2}{\Lambda^2_{\rm MS}}\right).
\end{equation}
As the energy scale is set by the temperature of the ensemble, one uses $\Lambda= 2\pi T$, and therefore the Free energy divided by $T^4$ scales as $\ln T$. If one computes the Pressure of this thermodynamic system, one gets:
\begin{equation}
    P= \frac{0.874017}{g^2}(4 K(-1))^3  \frac{3m a^4}{Na^3} T^4 = A_1 \ln(x) x^4 \frac{n}{(n-1)a^3},
\end{equation}
where $x=T/T_c$, $N$ is the total number of cubes in the discrete volume, and $A_1$ is a numerical constant which also includes the numerical fraction $a\sim r/T_c$, such that $aT\sim r T/T_c<<1$.

\begin{figure}[htbp]
    \centering
         \includegraphics[scale=0.5]{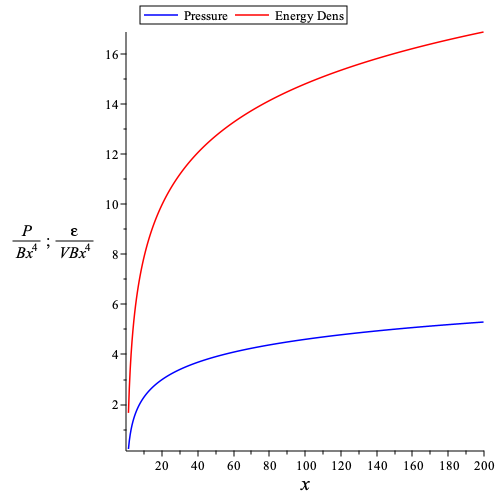}
    \caption{Plot of scaled pressure and energy with $x=T/T_c$. }
    \label{fig:presslat}
\end{figure}

Similarly, the energy  is found as 
\begin{equation}
\epsilon = T^2 \frac{\partial \ln Z}{\partial T} = m A_1 \ x^4 (1+ 3 \ln x).
\end{equation} 
In the following graph (\ref{fig:presslat}), we plot the graphs of $P/x^4$ and $\epsilon/V x^4$. In that case, the pre-factor of both the quantities is $B= A_1 n/(n-1) a^3$. 

Next, we add the fluctuations to the condensate and observe the change in the partition function. Unlike the above, the fluctuations' contribution to the partition function is numerically coded, and the output is plotted as a graph. As in thermal equilibrium, the background provides the leading order contribution to the thermodynamics of the system, but the fluctuations exchange energy with the surroundings/thermal bath. This would be the numerical lattice contribution, above the analytic background contribution to the Lattice partition function.

\subsection{Fluctuations}
\begin{equation}
    A_i^a= U(t) \delta^a_i + f_i \delta_{ij} \delta^a_j .
\end{equation}
where $f_i$ are three independent fields, which can be different for links, but given one plaquette in the $xz$ plane centered at a $(x,z)$, we take the fluctuations as constant over the links, making the 2-dimensional planar plaquette. These are therefore random numbers $(f_x,f_y,f_z)$, which might be the generators of the Metropolis code.
If we compute the plaquette contribution to the action on the $xz$ plane, centered at a point $(x,z)$, one has the trace as 
\begin{equation}
    {\rm Tr} (W_{xz}) = 2\left(1-2 \sin^2\left(\frac{U+f_x}{2}\right) \sin^2\left(\frac{U+f_z}{2}\right)\right).
\end{equation}

The action is therefore
\begin{equation}
    S(W_{xz})= -\frac{4}{g^2} \left(1- \frac12 {\rm Tr} (W_{xz})\right) .
\end{equation}

Using the series expansion of the hyperbolic and the trigonometric terms, one gets 
\begin{equation}
    S(W_{xz})=- \frac4{g^2} \left(\frac18 a^4 U_I^4 - \frac{a^4 U_I^2}{4} (f_x^2 + f_z^2) + \frac18 a^4 f_x^2 f_z^2\right).
\end{equation}
And in the above, we have kept the even powers of $U$ as otherwise it integrates to zero.

The combined action of the three different types of plaquettes, comprising three sides of a cube are (or sharing one vertex):
\begin{equation}
S(W_{xz},W_{xy},W_{yz})= -\frac1{g^2}\left(  \frac{3}{2} a^4 U_I^4 - 2 a^4 U_I^2 (f_x^2 +f_y^2 + f_z^2) + a^4 (f_x^2 f_z^2+ f_x^2 f_y^2 + f_y^2 f_z^2)\right).
\label{eqn:fluc}
\end{equation}

\begin{figure}[htbp]
    \centering
    \includegraphics[scale=0.5]{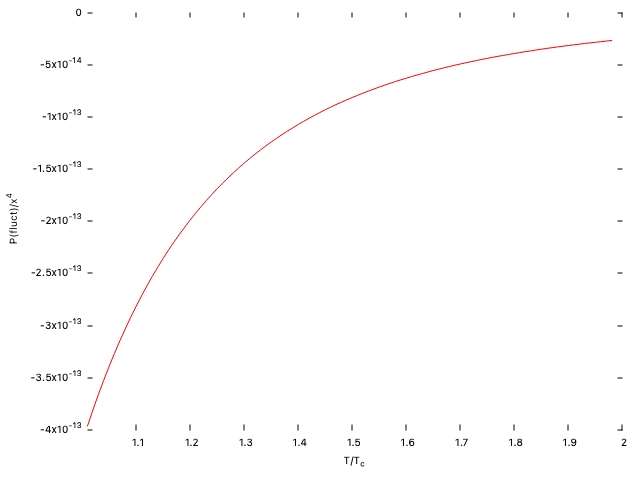}
    \caption{Contribution to Pressure from the fluctuations, a= $2.06 \times 10^{-9}$ in units of $1/T_c$ and 100 lattice sites.  }
    \label{fig:presslatfluct}
\end{figure}

The partition function for these would involve the sum over $f_x,f_y,f_z$ for all plaquettes. Using random numbers, one can fill up one configuration and then perform a sum over configurations. This calculation is rather cumbersome and takes a lot of computing power. 
We are showing the pressure obtained for `pure fluctuation' contribution in Fig. (\ref{fig:presslatfluct}), and that is negative due to the domination of the second term in Eq. (\ref{eqn:fluc}) in the calculation. However, if we add the contribution from the background, the overall pressure is positive again.
We have obtained the plots of the partition function, and similar features are seen as for the regular lattice predicted behaviour of the de-confined gluon gas. However, the number of configurations one integrates over is limited. We have work in progress as we require higher computing power to complete the calculations.

\section{Effect of Gravitational Waves on Thermodynamics of Quark-Gluon Plasma}{\label{sec:4}}

In this section, we study the effect of a GW on the thermodynamics of QGP. For this, we consider the GW-dependent condensate configuration explored in \cite{gnrcon}. In the paper \cite{gnrcon}, we studied the condensate + fluctuations (tensor decomposition) model in the presence of GWs. The gauge field configuration in the presence of GW  is given by 
\begin{equation}
   A^a_0 =0,\;  A^a_i = \delta^a_i U(t) + \delta^{aj} \tilde{A}_{ji} + \frac12 \delta^{aj}h_{ji} U(t),
\end{equation}
where $U(t)$ is the spatially homogeneous and isotropic condensate, $\tilde{A}_{ji}$ are the fluctuations without GWs and $h_{ji}$ is the $+$-polarised GW with only nonzero components, $h_{11} = -h_{22} = h_+ = A_+ \cos(\omega_g(t-z))$. Now consider an off-diagonal fluctuation to be non-zero, i.e. $\tilde{A}_{32}=\Phi_2(t)$. The generalisation of this case was investigated in great detail in \cite{gnrcon}, where we considered all different types of fluctuations over the condensate. The solutions for the equations of motion for $U(t)$ and $\Phi_2(t)$ are given in appendix \ref{app:sols}. It is obtained that $\Phi_2(t)$ can be in terms of $U(t)$ as 
\begin{equation}
    \Phi_2(t) = \frac{i}{\sqrt{3}} U(t).
\end{equation}
Then, the gauge field becomes
\begin{equation}
   A^a_i = \begin{pmatrix}
U +\frac12h_+ U & 0 & 0\\
0 & U-\frac12 h_+ U & 0\\
0 & \Phi_2 & U
\end{pmatrix}
\label{eqn:wgw}
\end{equation}

With the above gauge field, the form of thermodynamic potential is the same as in previous cases (Eq. \ref{eq:thdpot}) with a difference in the expressions of $F^{a\mu\nu} F^a_{\mu\nu}$ and $E^a_i E^{ai}$. 

\begin{multline}
    F^{a\mu\nu} F^a_{\mu\nu} = \frac{16}{3} U^4 -\frac23 h_+ U^4 -\frac{11}{2} h_+^2 U^4 +\frac23 h_+^3 U^4 + \frac{31}{24} h^4_+ U^4 -\frac18 h_+^6U^4 
    +\frac{16}{3} U'^2-\frac23 h_+ U'^2 - 3 h_+^2 U'^2 \\ - 2 h_+ U U' \partial_t h_+ + U^2 \left[ (\partial_z h_+)^2 + (\partial_t h_+)^2 \right],
    \label{eq:gwc}
\end{multline}
and 
\begin{equation}
    E^a_i E^{ai} = \frac{8}{3} U'^2 - \frac13 h_+ U'^2 - \frac32 h_+^2 U'^2 - h_+ U U' \partial_t h_+ + \frac12 U^2 (\partial_t h_+)^2.
\end{equation}

As a first approximation, we consider GW as a function of time only. Since the GWs are in thermal equilibrium with the condensate and fluctuations, we have to fix the parameters in the expressions of the condensate and GW in order to use the Matsubara frequency formalism. We did that with the condensate and fluctuation and found that the constant $c_1$ is related to temperature $T$ as $c_1 = 4\ i\ K(-1)\ T$. As for the GW expression, we have to expand the hyperbolic function in terms of the Matsubara frequency modes as
\begin{equation}{\label{eq:fourt}}
    \cosh(\omega_g \tau)= \sum_j a_j (\cos(2\pi j T  \tau) + b_j \sin (2 \pi j T \tau)).
\end{equation}
And the coefficients are found to be 
\begin{equation}
    a_j= 2T \int_0^{1/T} \cosh(\omega_g \tau) \cos(2  \pi  j T ~\tau) ~ d \tau=  \frac{T \omega_g}{4 j^2 \pi^2 T^2 +  \omega_g^2} \sinh\left(\frac{\omega_g}{T}\right),
    \label{eqn:fourt1}
\end{equation}
and 
\begin{equation}
    b_j= 2T \int_0^{1/T} \cosh(\omega_g \tau) \sin(2  \pi  j ~T ~\tau) ~ d \tau= - \frac{8T^2 ~j~ \pi}{4 j^2 \pi^2 T^2 + \omega_g^2} \sinh^2\left(\frac{\omega_g}{2T}\right).
    \label{eqn:fourt2}
\end{equation}

We keep only the terms upto $j=1$ in evaluating the partition function. Then, we end up with a thermodynamic potential depending on temperature $T$, constant $c_2$, GW amplitude $A_+$, and GW frequency $\omega_g$. For a comparison with lattice calculations, we plotted along $T/T_c$. 

\begin{figure}[htbp]
    \centering
    \includegraphics[width=0.5\linewidth]{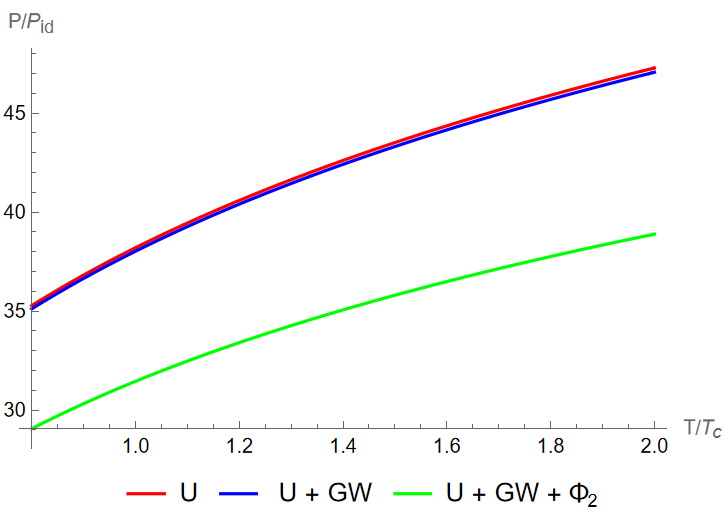}
    \caption{Normalised Pressure under different scenarios. We choose $A_+ = 0.4$ and $\omega_g = 10^{-2}$ for GW.}
    \label{fig:Pressall}
\end{figure}

\begin{figure}[htbp]
\centering
    \begin{subfigure}{0.4\textwidth}
    \centering
      \includegraphics[width=\textwidth]{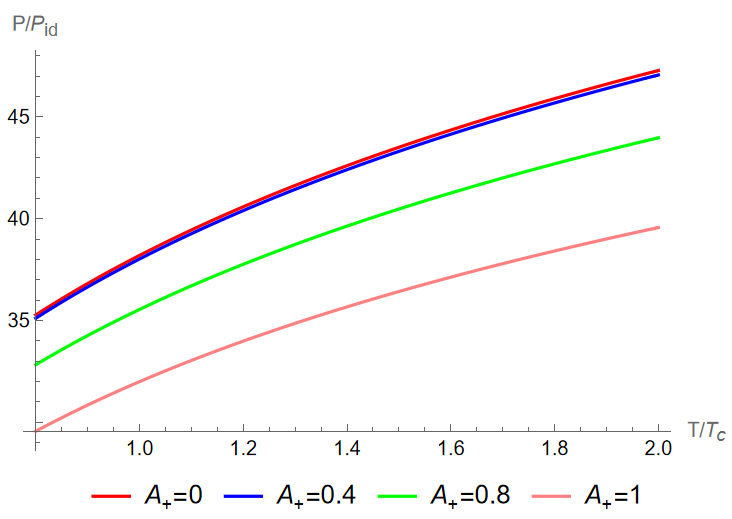}
      \caption{Normalized pressure for different GW  amplitudes ($\omega_g = 10^{-2}$).}
      \label{fig:PressdiffA}
    \end{subfigure}
    \hfill
    \begin{subfigure}{0.4\textwidth}
    \centering
      \includegraphics[width=\textwidth]{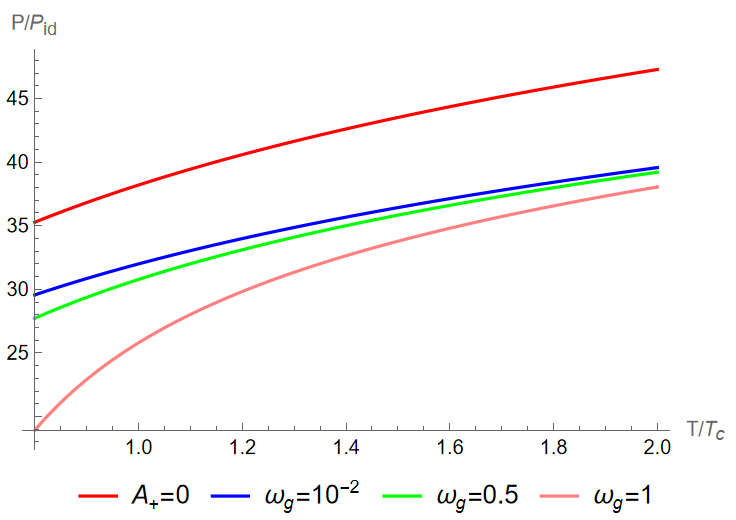}
      \caption{Normalized Pressures for different $\omega_g$ ($A_+ = 1$).}
      \label{fig:Pressdifffreq}
    \end{subfigure}  
    \caption{Normalised pressure as a function of temperature $T$ for different GW amplitudes and frequencies.}
    \label{fig:PressGWcase}
\end{figure}

Figs. \ref{fig:Pressall} and \ref{fig:PressGWcase} show the normalused pressure $P/P_{id}$ for the condensate background in the presence of GW perturbations. When either the fluctuation $\Phi_2$ or GW is included, the overall magnitude of the pressure decreases compared to the pure-condensate case. This behaviour can be understood from the structure of the classical action in the modified background.

The periodicity condition fixes the condensate amplitude such that the classical contribution to the effective action scales as $F^2_{\mu\nu} \sim U^4$, obtaining a very large pressure in the unperturbed case. The gauge field modifies anisotropically with the introduction of GWs as $A^1_1 = (1+ h_+/2) U$ and $A^2_2 = (1-h_+/2)U$ while the field $\Phi_2$ presents as an off-diagonal component. The off-diagonal element $\Phi_2 = i U/\sqrt{3}$ modifies the off-diagonal structure of the field strength tensor and partially cancels the contributions of the unperturbed case. As a result, the effective classical energy density decreases, and therefore, the pressure is reduced.

Increasing the GW amplitude or modifying the frequency enhances the deformation of the gauge field. Large values of $A_+$ increase the anisotropy between the spatial components of the gauge field, reducing the net classical energy density and leading to a monotonic reduction in pressure. This effect is clearly seen in Figs. \ref{fig:PressdiffA} and \ref{fig:Pressdifffreq}. This behaviour actually reflects the fact that the condensate background is not dynamically selected by minimizing the effective action. It is rather a high-energy density configuration that becomes less high when perturbed by either gauge fluctuations or metric perturbations. Thus, GW effectively lowers the background energy density by redistributing among field strength components and decreases their contribution to $F^2$. Even though the overall scale of the pressure remains much larger than lattice results, the inclusion of GWs produces a consistent and understandable reduction in the effective pressure.

\subsection{Adding GW to lattice}
To add GW to the condensate as fluctuations, one has to be careful, as there is a well-defined dependence on the $z$-coordinate. The GW as it arrives on the QGP has the form $h_+= A_+ \cos(\omega_g t- \omega_g z)$. As this gives rise to a different background, we recompute the QGP behaviour in a GW background at finite temperature. As the GW wave-function is hyperbolic in Euclidean time, we use a Fourier series of the function in terms of the Matsubara frequency modes. The $z= j a$, where $j$ is an integer, and $a$ is the lattice width. We also assume that the Euclidean time is periodic.
The gauge field in the presence of the GW is obtained as $A_{1}^1= U(1+ \frac12 h_+)$, $A_{2}^2= U(1-\frac12 h_+)$, the gauge field in the $z$-direction remains unchanged. 
Therefore, the contributions of the plaquette can be categorized as those of $W_{xz}$, $W_{yz}$ and $W_{yx}$. They can be computed as
\begin{equation}
    W_{xz}=e^{i a U(1+ \frac12 h_+(z)) \frac{\sigma^1}{2}}e^{ia U \frac{\sigma^3}{2}} e^{-ia U(1+ \frac12 h_+(z+a))\frac{\sigma^1}{2}}e^{-i a U \frac{\sigma^3}{2}}.
    \end{equation}
If we simplify the above and expand the hyperbolic and the trigonometric functions to linear order in the GW amplitude $A_+$, we get
\begin{equation}
    {\rm Tr} (W_{xz}) = 2\left[1- \frac1{8} U^4 a^4 (1+ A_+ \cosh(\omega_g \tau) \cos(\omega_g(j+1)a))\right].
\end{equation}
    A similar formula would follow for the $W_{yz}$ plaquette. 
The third type of plaquettes are those in the x-y plane, and 
\begin{equation}
    W_{xy}=e^{i a U(1+ \frac12 h_+(z)) \frac{ \sigma^1}{2}}e^{i a U(1- \frac12 h_+(z)) \frac{\sigma^2}{2}}e^{-i a U(1+ \frac12 h_+(z+a)) \frac{\sigma^1}{2}} e^{-i a U(1- \frac12 h_+(z+a)) \frac{\sigma^2}{2}}.
\end{equation}
What we find after simplification is that there is no linear contribution in $A_+$, as 
 \begin{eqnarray}
{\rm Tr} (W_{xy})&=& \left(\cos\left(\frac{U}4 (h_+ - h_{+a})\right)\cos\left(\frac U2(1-\frac12 h_+)\right)\cos\left(\frac U2(1-\frac12 h_{+a})\right) \right. \nonumber \\ &+& \left. \sin\left(\frac{U}2 (1-\frac12 h_+)\right)\sin\left(\frac{U}2 (1-\frac12 h_{+a})\right) \cos\left(U + \frac14 (h_+ + h_{+a})\right)\right).
\end{eqnarray}

The $h_+$ terms are quadratic in $A_+$ and higher due to the cosine functions. Thus, we have linear in GW contributions in plaquettes in the $x-z$ and the $y-z$ plane, and the ones in the $x-y$ plane; we ignore the GW contribution. However, the linear term $A_+$ in $W_{xz}$ is equal and opposite to the linear term in $W_{yz}$, and therefore, there is no significant contribution to the partition function from weak GW. The relevant terms would be proportional to $h_+^2$ as in the previous discussion in Eq. (\ref{eq:gwc}). However, introduction of a perturbation in $A^3_2 = \Phi_2(t) $  as motivated by \cite{gnrcon}, leads to a linear contribution to the Lattice partition function from the $h_+$. If we use the classical system with a fluctuation over the isotropic model of \cite{Prokhorov} as discussed in previous section and Appendix \ref{app:sols}, we find the solution of perturbation $\Phi_2(t)$ as $\Phi_2(t) = i \alpha U(t)$. 

If we add the GW to the equations, then the $\Phi_2$ equations are to linear order in $h_+$ and the transverse fields \cite{gnr}:
\begin{equation}
    -\Ddot\Phi_2 + \Phi_2 U^2 - h_+ \Phi_2 U^2 = 0.
\end{equation}
If we numerically investigate the solution for $\Phi_2$ with the GW, then the correction can be ignored for a short time. Therefore, we can ignore it in the computation of the Lattice action, where we are computing within $t\subset[0, \beta]$, which is $t\subset[0, 4K(-1)]$ in units of $\beta/4K(-1)$.  

Thus, if we sum over the three plaquette actions, which are three faces of the same cube with a vertex at $(x,y,z)$, one gets
\begin{equation}
S(W_{xy} W_{yz}, W_{xz})= - \frac{4}{g^2} \left(  \frac{3-\alpha^2}{8} a^4 U_I^4 + \frac{(\alpha^2)}4  U_I^4 a^4 A_+ \cosh(\omega_g \tau) \cos(\omega_g (j+1)a) \right),
\end{equation}
where $\alpha^2$ is the proportionality of the solution of $\Phi_2 = i \alpha U$.  
We then find the partition function using the Lattice action. However, the Fourier coefficients as well as the time integral are suppressed in powers of $1/j$; and therefore, in the computation of the partition function, we keep only the lowest term. Our results are also dependent on the $A_+$, and we do not integrate over all possible amplitudes of the GW. 
We then use the Fourier series for the hyperbolic function in terms of the Matsubara modes as in Eqs. (\ref{eqn:fourt1} \& \ref{eqn:fourt2}).

The integral of $\int U_I^4 \sin(2 \pi j \tau) d\tau=0$, the $\int U_I^4 \cos(4 j \pi \tau)~ d \tau $ is non-zero, but the coefficients fall off, and the integral also decreases as $n$ increases. Thus, we keep only the term $j = 1$ in the Action, and therefore, only the first Matsubara frequency mode is used in the evaluation of the partition function.

The integral 
\begin{equation}
    \int_0^{1/T} U_I^4 \cos(4 \pi T ~\tau)~ d\tau= -(4 K(-1))^3 T^3 (1.24769).
\end{equation}

Plugging this into the plaquette action, one sums over $(n-1)$ of the $z$ vertices, and multiplies them with $2n (n-1)$ to get the total contribution from this term to the plaquette action. A plot of the partition function and pressures shows almost the same behaviour of the thermodynamic systems as with the condensate due to the small amplitude of the GW. However, if we increase the amplitude of the GW to order $1$ or greater, a tangible effect appears on the thermodynamic quantities. 

{\small 
\begin{equation}
    S^{GW}_{\rm Lat}= \frac{ V (4 K(-1))^3 }{g^2} \left[ -3 m' \ A_1 x^3 +   2n(n-1)    A'_+ \left(-A_1\frac{x^4}{\tilde \omega_g} + A_2\frac{x^4 \tilde{\omega}_g}{16 \pi^2 x^2 + \tilde{\omega}_g^2} \right)\sinh\left(\frac{\tilde{\omega}_g}{x}\right)\sum_i \cos(\tilde{\omega}_g(i+1))\right].
\end{equation} }
where $m'= (3-\alpha^2)m/3$, $A_1=1.748038369$, $A_2=1.24769$, $\tilde{\omega}_g=\omega_g/T_c$, and $x=T/T_c$. We absorb the constant of proportionality that appears from $\Phi_2(t)$ in $A'_+$ and analyze the system accordingly.
The sum over $i$ can be computed analytically, and we get
\begin{equation}
\sum_{i=0}^{n-1} \cos((i+1)\tilde{\omega}_g)= \frac12 \left[\cos(\tilde{\omega}_gn) -1 + \frac{\sin(\tilde{\omega}_g)\sin(\tilde{\omega}_g n)}{1-\cos\tilde{\omega}_g}\right]=f(\tilde{\omega}_g,n).
\end{equation}

As this is an analytic formula for the total partition function, the pressure and energy density can be computed for the system. As the GW term is proportional to $A_+$, the contribution will be small, and the behaviour is nearly similar to the existence of the condensate by itself. 
We find 
\begin{equation}
    \Omega= - T S^{GW}_{\rm Lat}
\end{equation}
This can be written as
\begin{equation}
    \Omega= -\frac{V A x^4}{g^2}\left[1+ \frac{A'_+}{n-1}\left(\frac{x}{\tilde \omega_g} -\frac{A_2}{A_1}\frac{x \tilde{\omega}_g}{16 \pi^2 x^2 + \tilde{\omega}_g^2} \right)\sinh\left(\frac{\tilde{\omega}_g}{x}\right)f(\tilde\omega_g,n) \right].
\end{equation}
The Pressure from above is found as 
\begin{equation}
P= 2 A x^4 \beta_1 \ln(x)\left[1+ \frac{A'_+}{n-1}\left(\frac{x}{\tilde \omega_g} -\frac{A_2}{A_1}\frac{x \tilde{\omega}_g}{16 \pi^2 x^2 + \tilde{\omega}_g^2} \right)\sinh\left(\frac{\tilde{\omega}_g}{x}\right)f(\tilde\omega_g,n) \right].
\end{equation}

\begin{figure}[htbp]
\centering
\includegraphics[scale=0.5]{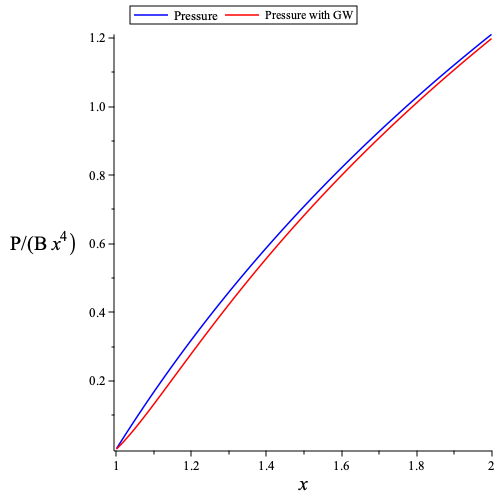}
\caption{Plots of Pressure without and with GW. The GW amplitude is $0.1$, and the frequency is $10$. The number of lattice sites is 100. }
\label{fig:freq}
\end{figure}

\begin{figure}[htbp]
\centering
\includegraphics[scale=0.5]{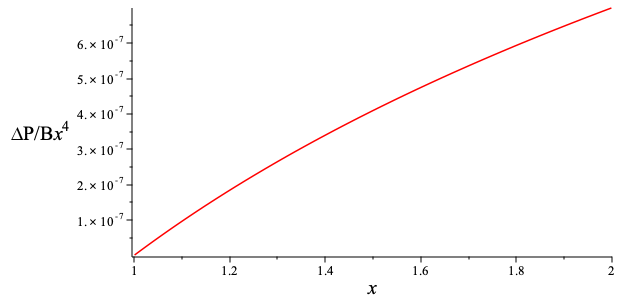}
\caption{Plots of Pressure with GW. The GW amplitude is $10^{-1}$ and frequency is $0.0001 T_c$. The number of lattice sites is 100. }
\label{fig:freq1}
\end{figure}

Observe that due to the $T_c \approx 100\ {\rm Mev}$, the GW frequency $\omega_g\approx 10^{20} {\rm Hz}$ in the graph of Fig. (\ref{fig:freq}). We plot the graphs for $\omega_g=0.0001 T_c$, and the pressure with the GW is almost the same as that without the GW. We have plotted the difference as in Fig. (\ref{fig:freq1}). For even lower frequencies, we replace the $\cos(\tilde{\omega}_g(i+1)) \approx 1,$ which gives an $n$ factor from the sum over Lattice sites.

\begin{figure}[htbp]
\centering
\includegraphics[scale=0.4]{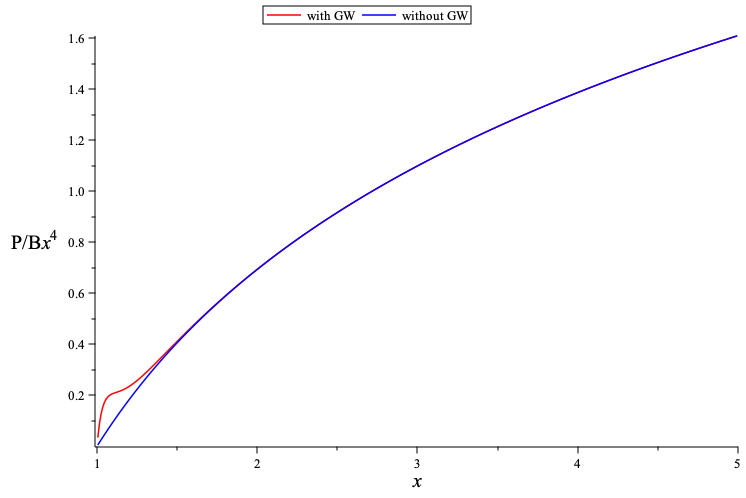}
\caption{Plot of pressure with GW at frequency $\omega_g= 20 T_c$, amplitude $A_+=0.001$, lattice sites 1000}
\label{fig:gw1}
\end{figure}

\begin{figure}[htbp]
\centering
\includegraphics[scale=0.4]{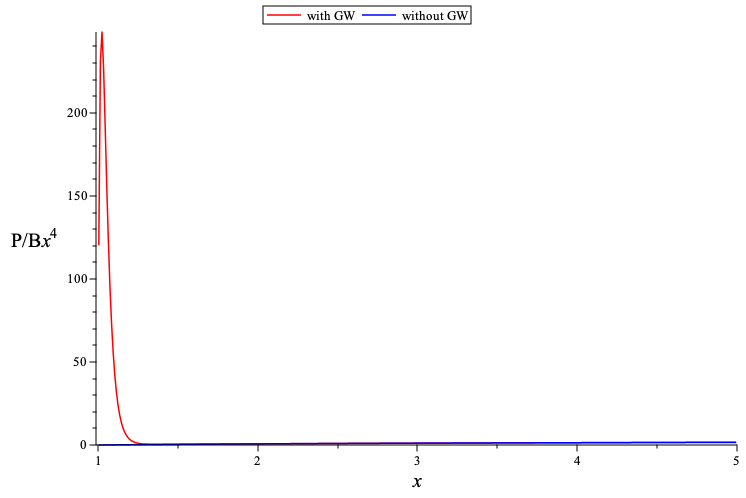}
\caption{Plot of pressure with GW at frequency $\omega_g= 46 T_c$, amplitude $A_+=10^{-10}$, lattice sites 1000}
\label{fig:gw2}
\end{figure}

 We observe, however, that due to the nature of the Fourier coefficient (\ref{eq:fourt}), there are particular frequencies for which the amplitude contribution increases (Figs. \ref{fig:gw1} \& \ref{fig:gw2}). At these frequencies, the behaviour of the pressure can fluctuate with temperature, and then stabilize back to that without the GW as $x$ increases. This sudden rise in pressure at a particular frequency is almost similar to a resonance behaviour, and we predict an instability in the gluon plasma at particular frequencies of the GW.

Due to the frequency dependence, if the GW has a high frequency, which it can have in the early universe, then the differences will be significant. If we worry about the number of Lattice sites, we can take the limit of lattice sites to infinity for this computation. We find that
\begin{equation}
    {\rm Lim}_{n\rightarrow \infty} \frac{f(\tilde{\omega}_g,n)}{n-1}=-1
\end{equation}
In this limit, the formula is independent of the lattice sites; only the volume is dependent on that. In that case, we find that for a given amplitude, there is a frequency for which the pressure formula shows instability as $x\rightarrow 1$. This is due to the hyperbolic functions, which appear in the formula, and they increase as $x/\tilde{\omega_g}$ decreases. Now, as we are using a GW solution for linearized gravity in Euclidean time, we have to restrict the hyperbolic function such that the $h_+ <<1$ approximation works for the Euclidean signature system too. For $x>>1$, the pressure behaviour is the same as without GW, and the actual magnitude is slightly perturbed by the GW. The existence of a `cutoff' frequency shows that to maintain equilibrium, we have to use non-perturbative solutions to the gravitational system. 

As these are obtained from $SU(2)$ partition functions, we refrain from making any predictions about QGP. However, certain features of the fluid suggest that one can have unexpected behaviour in a more detailed modelling with the $SU(3)$ group, which will actually have three $SU(2)$ subgroups and therefore three condensates \cite{gnr1}. 

The generic behaviour of the thermodynamics as derived here agrees with other studies of QGP using lattice methods \cite{borsanyi, latqgp}. Therefore, we think that our condensate model works well in the regime we are considering.

\section{Conclusions}

In this work, we investigated a type of time-dependent $SU(2)$ YM condensates as a non-perturbative classical background and explored their dynamical and thermodynamic implications. We first analysed the quarks in the condensate background and found that the backreaction can break the isotropy of the background for certain initial conditions. This shows that even a highly symmetric classical configuration can be changed by dynamical Fermionic degrees of freedom. However, if the Fermions have perturbative initial conditions, the thermodynamics of the system remains unchanged.

We then evaluated the one-loop finite temperature effective action using the background field method and heat-kernel expansion for the quark-gluon condensate model. The obtained thermodynamic quantities, such as pressure, capture the expected monotonic behavior of the pressure with temperature. However, the normalized pressure exhibits an approximately logarithmic temperature dependence rather than approaching the ideal gas limit, as seen in lattice gauge results. The magnitude of the pressure differs from the lattice results. This deviation originates from the dominant classical action density of the condensate, which scales as $T^4$ with a large prefactor fixed by the Euclidean periodicity condition. Therefore, our results highlight the conceptual distinction between semi-classical background- field analyses and ideal gas behavior. This is not unexpected for Plasma, as the conventional plasma pressure comprises of contribution from both the electrons and the interacting ions. We also compared our findings with the spectral approach of \cite{nonpert}, which constructs a thermal ensemble from a tower of massive excitations and successfully reproduces lattice results.  The differing behaviors of the two methods reflect their distinct physical assumptions: whereas the spectral model is designed to match equilibrium thermodynamics, the background-field approach probes the structure and energetics of non-trivial classical configurations. The discrepancy in pressure magnitudes thus arises naturally and does not indicate an inconsistency. The condensate model describes the QGP plasma state with a contribution from the free gas, as well as the gluon instanton. The quarks are in the de-confined phase, and therefore, the QGP we have studied is above the QGP phase transition temperature.

Next, we modelled the same system on a spatial lattice. The solution is a Euclidean instanton, and thus we keep the Euclidean time as periodic in inverse of temperature commensurate with finite temperature field theory. The spatial 3-dimensions are taken as a cubic lattice with lattice spacing as a fraction of the critical temperature length scale. The Wilson line and the partition function are calculated using the background condensate field and fluctuations. The Pressure behaves as found in the previous section if we restrict the condensate to be periodic in $\beta$, the inverse of the temperature. It grows as a log function of Temperature, and the condensate, which is an instanton, contributes a pressure which exceeds the ideal gas non-interacting pressure. The addition of fluctuations along the lattice directions decreases the pressure, but as these are non-interacting and random, their contribution is significantly less than the instanton contribution.

We further examined the influence of GW perturbations on the condensate by introducing a time-dependent anisotropy in the gauge fields and a non-diagonal component $\Phi_2$. The field $\Phi_2$ modifies the field-strength tensor in a way that reduces the classical action density. As a result, the pressure decreases when gravitational waves or gauge fluctuations are included, and the magnitude decreases further with increasing gravitational-wave amplitude or changes in frequency. This shows that the high-pressure behavior of the unperturbed condensate is sensitive to perturbations that might extract energy from the condensate. This is commensurate with the observations of our previous paper, which had observed a decay of the condensate to its plasmons \cite{gnrcon}. In this paper, we show that the plasmons decrease the pressure of the QGP. 

Adding GW to the lattice model and expanding the GW in Matsubara frequencies shows that the pressure of the QGP decreases. As the time is Euclideanized, the above is a stochastic GW in equilibrium with a finite temperature Yang-Mills plasma. To see a tangible interaction of the GW with the condensate, we had to excite one of the plasmon modes obtained in \cite{gnrcon}. At certain GW frequencies, we observe instabilities in the behavior of the pressure, and these could be due to (i) Plasmons being generated due to `resonance', (ii) Temperature equilibrium introduces a frequency cutoff of the GW background. The instability stabilizes with an increase in temperature, and they occur near the critical temperature. These, if found real in an experiment, the GW could induce the hadronization and phase transition; however, much work is required to verify these results. 

Our analysis clarifies the role and limitations of time-dependent YM condensates in semi-classical studies of non-Abelian gauge theories at finite temperature. While such configurations offer valuable insight into possible non-perturbative structures, much work is required to understand the behavior near the critical temperature. The present study opens several directions for further investigation. A natural extension is to examine the dynamical stability of the time-dependent condensate by including higher-order quantum corrections or alternative fluctuation modes. This may reveal nearby classical configurations with lower action densities and provide insight into whether modified condensates could yield more realistic thermodynamic behavior.

Another promising direction is to explore different non-linear Yang–Mills backgrounds that are not exactly fixed by the Euclidean periodicity condition, since such configurations might avoid the large classical contribution responsible for the enhanced pressure found here. Likewise, incorporating gravitational waves more dynamically could clarify gauge–gravity interactions in early-universe settings. Finally, the comparison with spectral approaches suggests investigating hybrid frameworks that combine semi-classical backgrounds with dynamically generated excitation spectra. Such models may help bridge the gap between background-field methods and phenomenological descriptions that successfully match lattice thermodynamics. These avenues offer a path toward a richer understanding of non-perturbative gauge dynamics at finite temperature. We also need to incorporate the dynamics of the Polyakov loop to understand confinement, and include quarks in lattice calculations to comprehend chiral phase transitions.

\appendix

\section{Evaluation of Effective action}{\label{app:1}}
\subsection{Gluon Contribution}
In this section, we will find the contribution of gluons to the one-loop effective action. The gluon contribution to the effective action is 
\begin{equation}
    \Gamma_g =\frac{1}{2} {\rm Tr} \ln \left(-\hat{\nabla}^2 \delta_{\mu\nu} - 2 \hat{F}_{\mu\nu} \right) - {\rm Tr} \ln (-\hat{\nabla}^2)
\end{equation}
Applying the Heat kernel expansion procedure, we get 
\begin{equation}
    \Gamma_g = -\frac{1}{2} \Lambda^{2\epsilon} \int_0^\infty \frac{d\tau}{\tau} \frac{1}{(4\pi \tau)^{d/2}} \int d^dx \sum_{n=0}^\infty \tau^n b_{n,g}^T(x) := \int d^dx \mathcal{L}_g,
\end{equation}
where 
\begin{equation}
    \mathcal{L}_g = -\frac{1}{2} \Lambda^{2\epsilon} \int_0^\infty \frac{d\tau}{\tau} \frac{1}{(4\pi \tau)^{d/2}} \sum_{n=0}^\infty \tau^n b_{n,g}^T(x)
\end{equation}
with $b^T_{n,g}=b^T_{n,gl}-2 b^T_{n,gh}$ are Seeley-deWitt coefficients, the first one ($b^T_{n,gl}$) corresponds to the gluon operator and the second one ($b^T_{n,gh}$) corresponds to the ghost operator. We use dimensional regularization to regulate ultraviolet divergences by choosing the convention $d=4-2\epsilon$ and introducing a mass renormalization scale $\Lambda^{2\epsilon}$.

The explicit expressions for the coefficients are given by 
\begin{align}
    b_{0,g}^T & = (d-2)\; {\rm tr_c}(\varphi_0),\\
    b^T_{1,g} & = 0,\\
    b^T_{2,g} & = \left(-2+\frac{d-2}{12}\right) {\rm tr_c}(-\varphi_0\; \hat{F}_{\mu\nu}^2) - \frac{d-2}{6} {\rm tr_c} (-\bar{\varphi}_2\; \hat{E}_i^2),
\end{align}
In the above, $\varphi_n(L)$ corresponds to bosonic versions ($\varphi_n^+(L)$) and all the terms are in the adjoint representation. 

Let us rewrite the gluon effective Lagrangian as follows
\begin{equation}
    \mathcal{L}_g = \mathcal{L}_{0,g}+\mathcal{L}_{1,g}+\mathcal{L}_{2,g}+ ...,
\end{equation}
where $\mathcal{L}_{0,g},\mathcal{L}_{1,g},\mathcal{L}_{2,g},...$ corresponds to $b^T_{0,g},b^T_{1,g},b^T_{2,g},..$, respectively. While evaluating the functional determinants, we encounter the following type of integrals
\begin{equation}
    I^+_{n,\alpha} = \int_0^\infty d\tau \tau^\epsilon \tau^{\alpha-1} \varphi_n^+ .
\end{equation}
These integrals can be done in a closed form (See Appendix \ref{app:2}). We get 
\begin{equation}
    I^+_{n,\alpha}  = 4 \pi^{1/2} i^n \frac1\beta \left(\frac{\beta}{2\pi}\right)^{2\epsilon} \left(\frac{\beta}{2\pi}\right)^{2\alpha+1} \Gamma\left(\alpha+\epsilon+\frac{n}{2} +\frac12\right) \zeta(2\alpha+2\epsilon+1),
\end{equation}
where $\Gamma(z)$ is the Gamma function and $\zeta(s)$ is the Riemann Zeta function. Now, we will find each term's contribution to the effective Lagrangian. The zeroth order term requires only one integral $I^+_{0,-2}$.
\begin{align}
    \mathcal{L}_{0,g} & = -\frac12 \Lambda^{2\epsilon} \int_0^\infty \frac{d\tau}{\tau} \frac{1}{(4\pi\tau)^{d/2}} b^T_{0,g}\\
    & = - \frac{1}{(4\pi)^2} (N^2-1) (4\pi\Lambda^2)^\epsilon \int_0^\infty d\tau \tau^{\epsilon-3} \varphi^+_0\\
    & = - \frac{1}{(4\pi)^2} (N^2-1) (4\pi\Lambda^2)^\epsilon I^+_{0,-2}\\
    & = - \frac1{4 \pi^{3/2}\beta}   (N^2-1) (4\pi\Lambda^2)^\epsilon \left(\frac{\beta}{2\pi}\right)^{2\epsilon}\left(\frac{\beta}{2\pi}\right)^{-3} \Gamma\left(\epsilon-\frac32\right) \zeta(2\epsilon-3)\\
    & = - \frac{\pi^2}{45} T^4 (N^2-1),
\end{align}
where we used the following expansions: $\Gamma (\epsilon-3/2) \approx 4\sqrt{\pi}/3 + \mathcal{O}(\epsilon)$ and $\zeta(2\epsilon-3)\approx 1/{120}+\mathcal{O}(\epsilon) $ as $\epsilon \rightarrow 0$.
Similarly, we can find the second-order term's contribution as follows
\begin{equation}
    \mathcal{L}_{2,g}  = -\frac12 \Lambda^{2\epsilon} \int_0^\infty \frac{d\tau}{\tau} \frac{1}{(4\pi\tau)^{d/2}} \tau^2 b^T_{2,g}.
\end{equation}
Using the $b^T_{2,g}$ and integrals notation ($I^+_{n,\alpha}$), we can evaluate as follows
\begin{align}
    \mathcal{L}_{2,g} & = \frac{1}{(4\pi)^2} \frac{11}{12} (4\pi\Lambda^2)^\epsilon I^+_{0,0} {\rm tr_c}(-\hat{F}_{\mu\nu}^2) + \frac{1}{(4\pi)^2} \frac{1}{6} (4\pi\Lambda^2)^\epsilon (I^+_{0,0}+2I^+_{2,0}) {\rm tr_c} (-\hat{E}_i^2)\\
    & = \frac{1}{(4\pi)^2} \frac{11}{12} 2 \pi^{-1/2} \left[ (4\pi\Lambda^2)^\epsilon \left(\frac{\beta}{2\pi}\right)^{2\epsilon} \Gamma\left(\epsilon+\frac12\right) \zeta(2\epsilon+1) \right] {\rm tr_c} (-\hat{F}^2_{\mu\nu}) \nonumber \\
     & + \frac{1}{(4\pi)^2} \frac{1}{6} 2 \pi^{-1/2} \left[ (4\pi\Lambda^2)^\epsilon \left(\frac{\beta}{2\pi}\right)^{2\epsilon} \zeta(2\epsilon+1) \left(\Gamma\left(\epsilon+\frac12\right) - 2\Gamma\left(\epsilon+\frac32\right)\right) \right] {\rm tr_c}(-\hat{E}_i^2)\\
     & = \frac{1}{(4\pi)^2} \frac{11}{12} \left[\frac1\epsilon +\gamma_E + \ln(4\pi) \right] {\rm tr_c} (-\hat{F}^2_{\mu\nu}) +\frac{11}{6} \frac{1}{(4\pi)^2} \ln\left( \frac{\Lambda}{4\pi T}\right) {\rm tr_c} (-\hat{F}^2_{\mu\nu}) -\frac{1}{3} \frac{1}{(4\pi)^2} {\rm tr_c}(-\hat{E}^2_i).
\end{align}
One can see that the first term has a pole at $\epsilon=0$ and the divergent terms can be removed by adopting $\Bar{MS}$ scheme. We will discuss this in the next section along with quarks.

Finally, combining all terms up to mass dimension 4, we get 
\begin{align}
    \mathcal{L}_g &= \mathcal{L}_{0,g} + \mathcal{L}_{0,g} + \mathcal{L}_{0,g}\\
    & = - \frac{\pi^2}{45} T^4 (N^2-1)+ \frac{1}{(4\pi)^2} \frac{11}{12} \left[\frac1\epsilon +\gamma_E + \ln(4\pi) \right] {\rm tr_c} (-\hat{F}^2_{\mu\nu}) +\frac{11}{6} \frac{1}{(4\pi)^2} \ln\left( \frac{\Lambda}{4\pi T}\right) {\rm tr_c} (-\hat{F}^2_{\mu\nu}) \nonumber\\
    & -\frac{1}{3} \frac{1}{(4\pi)^2} {\rm tr_c}(-\hat{E}^2_i).
\end{align}

\subsection{Quark Contribution}
In this section, we will study the contribution of quarks to the one-loop effective action and on thermodynamic potential. The quark contribution is given by 
\begin{equation}
    \Gamma_q = -N_f {\rm Tr} \ln (\slashed \nabla),\; \; \slashed \nabla= \gamma_\mu \nabla_\mu.
\end{equation}
To apply the heat kernel procedure, we need to find the Klein-Gordon operator related to the above one. One can use the following manipulations:
\begin{equation}
    {\rm Tr}\ln (\slashed \nabla) = \frac12 {\rm Tr} \ln (-\slashed \nabla^2),\; \; -\slashed \nabla^2 = -\nabla^2-\frac12 \sigma_{\mu\nu} F_{\mu\nu},\; \; \sigma_{\mu\nu} = \frac12 [\gamma_\mu,\gamma_\nu],
\end{equation}
which makes the quark effective action as 
\begin{equation}
    \Gamma_q = - \frac12 {\rm Tr} \ln \left( -\nabla^2-\frac12 \sigma_{\mu\nu} F_{\mu\nu} \right).
\end{equation}
Now, again using Heat kernel expansion, we get 
\begin{equation}
    \Gamma_q = \frac{N_f}{2} \Lambda^{2\epsilon} \int_0^\infty \frac{d\tau}{\tau} \frac{1}{(4\pi \tau)^{d/2}} \int d^dx \sum_{n=0}^\infty \tau^n b_{n,q}^T(x) := \int d^dx \mathcal{L}_q,
\end{equation}
where 
\begin{equation}
    \mathcal{L}_q = \frac{N_f}{2} \Lambda^{2\epsilon} \int_0^\infty \frac{d\tau}{\tau} \frac{1}{(4\pi \tau)^{d/2}} \sum_{n=0}^\infty \tau^n b_{n,q}^T(x)
\end{equation}
with $b^T_{n,q}$ being the Seeley-deWitt coefficients which are given by 
\begin{align}
    b^T_{0,q} & = 4\; {\rm tr_c}(\varphi^-_0),\\
    b^T_{1,q} & = 0,\\
    b^T_{2,q} & = -\frac23 {\rm tr_c} (-\varphi^-_0 F^2_{\mu\nu}) - \frac23 {\rm tr_c}(-\bar{\varphi}^-_2 E^2_i), 
\end{align}
where $\varphi_n^-(L)$ is the fermionic version and all the terms are in the fundamental representation. Similarly, let us rewrite the quark-effective Lagrangian as
\begin{equation}
    \mathcal{L}_q = \mathcal{L}_{0,q} + \mathcal{L}_{1,q} + \mathcal{L}_{2,q} + ...,
\end{equation}
where $\mathcal{L}_{0,q},\mathcal{L}_{1,q},\mathcal{L}_{2,q},...$ corresponds to the coefficients $b^T_{0,q},b^T_{1,q},b^T_{2,q},..$, respectively. We can proceed to evaluate the terms individually. We note that we encounter the same type of integrals as in the gluon case [See Appendix \ref{app:2}]:
\begin{equation}
    I^-_{n,\alpha} = \int_0^\infty d\tau \tau^\epsilon \tau^{\alpha-1} \varphi_n^- = 4 \pi^{1/2} i^n \frac1\beta \left(\frac{\beta}{2\pi}\right)^{2\epsilon} \left(\frac{\beta}{2\pi}\right)^{2\alpha+1} \Gamma\left(\alpha+\epsilon+\frac{n}{2} +\frac12\right) \zeta\left(2\alpha+2\epsilon+1,\frac12\right),
\end{equation}
where $\zeta(a,z)=\Sigma_{n=0}^\infty (n+z)^{-a}$ is the generalized Riemann-zeta function. Now, the zeroth order term of effective Lagrangian is 
\begin{align}
    \mathcal{L}_{0,q} & =  \frac{N_f}{2} \Lambda^{2\epsilon} \int_0^\infty \frac{d\tau}{\tau} \frac{1}{(4\pi \tau)^{d/2}} b_{0,q}^T(x)\\
    & = 2 N N_f \Lambda^{2\epsilon} \int_0^\infty \frac{d\tau}{\tau} \frac{1}{(4\pi\tau)^{d/2}} \varphi^-_0\\
    & = \frac{2}{(4\pi)^2} N N_f (4\pi\Lambda^2)^\epsilon I^-_{0,-2}\\
    & = 4 N N_f T^4 \pi^{3/2} (4\pi\Lambda^2)^\epsilon \left(\frac{\beta}{2\pi} \right)^{2\epsilon} \Gamma\left(\epsilon-\frac32\right)\zeta\left(2\epsilon-3,\frac12\right)\\
    & = - \frac{7\pi^2}{180} N N_f T^4,
\end{align}
where we used the expansions of Gamma and generalized Riemann zeta functions as $\epsilon \rightarrow 0$. Similarly, we can find the second order's contribution 
\begin{align}
     \mathcal{L}_{2,q} &  = \frac{N_f}{2} \Lambda^{2\epsilon} \int_0^\infty \frac{d\tau}{\tau} \frac{1}{(4\pi\tau)^{d/2}} \tau^2 b^T_{2,q}\\
     & = -\frac{1}{(4\pi)^2} \frac{N_f}{3} (4\pi\Lambda^2)^\epsilon I^-_{0,0} {\rm tr_c}(-F_{\mu\nu}^2) - \frac{1}{(4\pi)^2} \frac{N_f}{3} (4\pi\Lambda^2)^\epsilon (I^-_{0,0}+2I^-_{2,0}) {\rm tr_c} (-E_i^2)\\
    & = - \frac{1}{(4\pi)^2} \frac{N_f}{3} 2 \pi^{-1/2} \left[ (4\pi\Lambda^2)^\epsilon \left(\frac{\beta}{2\pi}\right)^{2\epsilon} \Gamma\left(\epsilon+\frac12\right) \zeta\left(2\epsilon+1,\frac12\right) \right] {\rm tr_c} (-F^2_{\mu\nu}) \nonumber \\
     & - \frac{1}{(4\pi)^2} \frac{N_f}{3} 2 \pi^{-1/2} \left[ (4\pi\Lambda^2)^\epsilon \left(\frac{\beta}{2\pi}\right)^{2\epsilon} \zeta\left(2\epsilon+1,\frac12\right) \left(\Gamma\left(\epsilon+\frac12\right) - 2\Gamma\left(\epsilon+\frac32\right)\right) \right] {\rm tr_c}(-E_i^2)\\
     & = -\frac{1}{(4\pi)^2} \frac{N_f}{3} \left[\frac1\epsilon +\gamma_E + \ln(4\pi) \right] {\rm tr_c} (-F^2_{\mu\nu}) -\frac{N_f}{3} \frac{1}{(4\pi)^2} \left[2 \ln\left( \frac{\Lambda}{4\pi T}\right) + 2 \ln4 \right] {\rm tr_c} (-F^2_{\mu\nu}) \\
     & +\frac{1}{(4\pi)^2} \frac{2}{3} N_f  {\rm tr_c}(-E^2_i).
\end{align}
Here also, we can see that the first term has a pole as $\epsilon=0$. Combining all the terms in quark contribution (up to mass dimension 4), we get 
\begin{multline}
    \mathcal{L}_q = -\frac{7\pi^2}{180} N N_f T^4 +\frac{1}{(4\pi)^2} \frac{N_f}{3} \left[\frac1\epsilon +\gamma_E + \ln(4\pi) \right] {\rm tr_c} (F^2_{\mu\nu}) + \frac{N_f}{3} \frac{1}{(4\pi)^2} \left[2 \ln\left( \frac{\Lambda}{4\pi T}\right) + 2 \ln4 \right] {\rm tr_c} (F^2_{\mu\nu})\\
    - \frac{1}{(4\pi)^2} \frac{2}{3} N_f  {\rm tr_c}(E^2_i).
\end{multline}    

\subsection{Renormalised Lagrangian}

Finally, we consider the treatment for the divergent terms in both gluon and quark effective actions. Consider the divergent terms in quark and gluon along with the tree Lagrangian:
\begin{multline}
    \mathcal{L}_{0} + \mathcal{L}_g^{div} + \mathcal{L}_q^{div} = \frac{1}{4g_0^2} F^{a\mu\nu} F^a_{\mu\nu} -  \frac{1}{(4\pi)^2} \frac{11}{12} \left[\frac1\epsilon +\gamma_E + \ln(4\pi) \right] {\rm tr_c} (\hat{F}^2_{\mu\nu}) \\
    + \frac{1}{(4\pi)^2} \frac{N_f}{3} \left[\frac1\epsilon +\gamma_E + \ln(4\pi) \right] {\rm tr_c} (F^2_{\mu\nu}).
\end{multline}
Using the trace properties of SU(N) matrices (${\rm tr_c}(\hat{F}^2_{\mu\nu}) = 2 N \; {\rm tr_c}(F^2_{\mu\nu}) = N F^{a\mu\nu} F^a_{\mu\nu} $), we get
\begin{equation}
    \mathcal{L}_{0} + \mathcal{L}_g^{div} + \mathcal{L}_q^{div} = \frac{1}{4g^2(\Lambda)} F^{a\mu\nu} F^a_{\mu\nu},
\end{equation}
with the standard one-loop beta function using $\Bar{MS}$ scheme,
\begin{equation}
    \frac{1}{g^2(\Lambda)} = \frac{1}{g^2_0} - \beta_1 \left( \frac1\epsilon + \ln(4\pi) + \gamma_E \right),\; \; \; \beta_1 = \frac{1}{(4\pi)^2} \left( \frac{11}{3} N -\frac23 N_f \right).
\end{equation}

Finally, putting together all the terms (both gluon and quark) up to mass dimension 4, we find 
\begin{multline}
    \mathcal{L} = - \frac{\pi^2}{45} T^4 (N^2-1)-\frac{7\pi^2}{180} N N_f T^4 +\left[\frac{1}{4g^2(\Lambda)} - \frac12 \beta_1 \ln\left(\frac{\Lambda}{4\pi T}\right) + \frac{1}{(4\pi)^2} \frac{N_f}{3} \ln 4 \right] F^{a\mu\nu} F^a_{\mu\nu} \\
    - \frac13 \frac{1}{(4\pi)^2} (N_f-N) E^{ai} E^a_i
\end{multline}

\section{Integrals used in Appendix A}{\label{app:2}}
The basic integrals are of the form
\begin{equation}
     I^\pm_{n,\alpha} = \int_0^\infty d\tau \tau^\epsilon \tau^{\alpha-1} \varphi_n^\pm (L)
\end{equation}
where the function $\varphi_n^\pm (L)$ are defined by Eq. \ref{eq:ploop} and $\pm$ corresponds to bosonic and fermionic versions, respectively. For the bosonic version, the integrals are of form
\begin{equation}
     I^+_{n,\alpha} = \int_0^\infty d\tau \tau^\epsilon \tau^{\alpha-1} \varphi_n^+ (L)
\end{equation}
where $\varphi_n^+ (L)$ is given by 
\begin{equation}
        \varphi_n^+(L) = (4\pi\tau)^{1/2} \frac{1}{\beta} \sum_{p_0^+} \tau^{n/2} R^n e^{\tau R^2}, \; \; R=i p_0^+ -\frac{1}{\beta} \ln(L), \; \;p_0^+ = \frac{2\pi k}{\beta}
\end{equation}
Using the condensate ansatz ($A^a_0=0,A^a_i = \delta^a_i U $), one can find the $\varphi_n^+ (L)$ is
\begin{equation}
    \varphi_n^+(L) = (4\pi\tau)^{1/2} \frac{1}{\beta} \left(\frac{2\pi i}{\beta}\right)^n \sum_{k\in \mathbb{Z}} \tau^{n/2} k^n e^{-\frac{4\pi^2 k^2}{\beta^2}\tau}
\end{equation}
Using this, we evaluate the integral as 
\begin{align}
    I^+_{n,\alpha} &= \int_0^\infty d\tau \tau^\epsilon \tau^{\alpha-1} \varphi_n^+ (L)\\
    & = \int_0^\infty d\tau \tau^\epsilon \tau^{\alpha-1}  (4\pi\tau)^{1/2} \frac{1}{\beta} \left(\frac{2\pi i}{\beta}\right)^n \sum_{k\in \mathbb{Z}} \tau^{n/2} k^n e^{-\frac{4\pi^2 k^2}{\beta^2}\tau}\\
    & = \frac{1}{\beta} \left(\frac{2\pi i}{\beta}\right)^n \sqrt{4\pi} \sum_{k \in \mathbb{Z}} k^n \int_0^\infty d\tau \tau^{\alpha+\epsilon+\frac{n}{2}+\frac12-1} e^{-\frac{4\pi^2 k^2}{\beta^2}\tau}\\
    & = i^n \frac{\sqrt{4\pi}}{\beta} \left(\frac{\beta}{2\pi}\right)^{2\alpha+1+2\epsilon} \Gamma\left(\alpha+\epsilon+\frac{n}{2}+\frac12\right) \sum_{k\in \mathbb{Z}} \frac{1}{k^{2\alpha+2\epsilon+1}}\\
    & = 2 i^n \frac{\sqrt{4\pi}}{\beta} \left(\frac{\beta}{2\pi}\right)^{2\alpha+1+2\epsilon} \Gamma\left(\alpha+\epsilon+\frac{n}{2}+\frac12\right) \zeta(2\alpha+2\epsilon+1)
\end{align}
where we used the definitions of Gamma and Riemann zeta functions \cite{Gradshteyn}: $\Gamma(z) = \int_0^\infty dt\; t^{z-1} e^{-t} $ and $\zeta(z) = \sum_{n=0}^\infty 1/n^z$.

For the Fermionic version, the integrals are form 
\begin{equation}
     I^-_{n,\alpha} = \int_0^\infty d\tau \tau^\epsilon \tau^{\alpha-1} \varphi_n^- (L)
\end{equation}
where $\varphi_n^- (L)$ is given by 
\begin{equation}
        \varphi_n^-(L) = (4\pi\tau)^{1/2} \frac{1}{\beta} \sum_{p_0^-} \tau^{n/2} R^n e^{\tau R^2}, \; \; R=i p_0^- -\frac{1}{\beta} \ln(L), \; \;p_0^- = \frac{2\pi}{\beta}(k+\frac12),
\end{equation}
As we can see the only difference from the bosonic versions is $k$ is replaced by $k+\frac12$. In the same way, one can evaluate the integral and the final expression is 
\begin{equation}
    I^-_{n,\alpha} = 2 i^n \frac{\sqrt{4\pi}}{\beta} \left(\frac{\beta}{2\pi}\right)^{2\alpha+1+2\epsilon} \Gamma\left(\alpha+\epsilon+\frac{n}{2}+\frac12\right) \zeta\left(2\alpha+2\epsilon+1,\frac12\right)
\end{equation}
where we used the definition of the generalized Riemann zeta function \cite{Gradshteyn}: $\zeta(z,a) = \sum_{n=0}^\infty 1/(n+a)^z$.



\section{Calculation of Wilson Action}{\label{app:wilson}}
The Calculation of the Wilson action: 
Without the GW, the action for a plaquette with sides along the $x-y$ direction can be calculated as:
\begin{equation}
W_{xy} = \exp\left(i a U(t) \frac{\sigma^1}{2}\right)\exp\left(i a U(t) \frac{\sigma^2}{2}\right) \exp\left(- i a U(t) \frac{\sigma^1}{2}\right) \exp\left(-i a U(t) \frac{\sigma^2}{2}\right)
\end{equation}
Using the formula for exponentiated matrices, one gets:
\begin{multline}
=\left(\cos\left(\frac{aU}{2}\right) + i \sigma^1\sin\left(\frac{aU}{2}\right)\right)\left(\cos\left(\frac{aU}{2}\right) + i \sigma^2\sin\left(\frac{aU}{2}\right)\right)\\
\left(\cos\left(\frac{aU}{2}\right) - i \sigma^1\sin\left(\frac{aU}{2}\right)\right)\left(\cos\left(\frac{aU}{2}\right) - i \sigma^2\sin\left(\frac{aU}{2}\right)\right)
\end{multline}
Out of all the terms in the above expression, only those which contribute to the trace are identified:
\begin{equation}
    = \left(\cos^4 \left(\frac{aU}{2}\right)- \sin^4 \left(\frac{aU}{2}\right) + 2 \cos^2 \left(\frac{aU}{2}\right) \sin^2 \left(\frac{aU}{2}\right)\right) I
\end{equation}
This can be further simplified to 
\begin{equation}
W_{xy}= \left(1- 2 \sin^4 \left(\frac{aU}{2}\right)\right) I.
\end{equation}
The other plaquette contribution for the Isotropic systems would thus have the same contribution, and the Wilson action is:
\begin{equation}
S_{xy}= - \frac{4}{g^2}\int_0^{\beta}  \left( 1 - {\rm Tr} \ W_{xy}\right)= -\frac{8}{g^2} \int_0^\beta \sin^4 \left(\frac{aU}{2}\right) d\tau
\end{equation}
where we are integrating over Euclidean time. If we use the Euclidean time in the condensate, i.e. $U(t)\rightarrow U(-i \tau)$, one gets using the property of the Lemniscate function 
$sin(U)\rightarrow i \sinh (U_I)$.
On adding the GW and the longitudinal mode as in \cite{gnrcon}, one gets 
\begin{equation}
    W_{yz}= \exp\left(i a \left(U_h\frac{\sigma^2}{2} + \Phi_2 \frac{\sigma^3}{2}\right)\right) \exp\left(i a U \frac{\sigma^3}{2}\right)\exp\left( -i a \left(U_{h+a}\frac{\sigma^2}{2} + \Phi_2 \frac{\sigma^3}{2}\right)\right)\exp\left(-i a U \frac{\sigma^3}{2}\right)
\end{equation}
where $U_h= U(1- \frac12 h_+ (t,z))$, and $U_{h+a}= U (1- \frac12 h_+(t,z+a))$. Further, using $\bar U_h = a\sqrt{ U_h^2 + \Phi_2^2}, \ \  \bar U_{h+a}= a\sqrt{U_{h+a}^2 + \Phi_2^2}$, one gets from the above
\begin{equation}
    W_{yz}= \left(\cos \frac{\bar U_h}2 \cos \frac{\bar U_{h+a}}2 + a^2 \frac{U_h U_{h+a} \cos(aU) + \Phi_2 ^2}{\bar U_h \bar U_{h+a}}~ \sin \frac{\bar U_h}2 \sin\frac{\bar U_{h+a}}2 \right)I 
\end{equation}
However, to the first order, there is no contribution from the $\Phi_2$ to the plaquette if we focus on the scalar algebraic terms,
\begin{equation}
    W_{yz}= \cos \left(\frac{\bar{U}_{h+a}-\bar{U_{h}}}{2}\right) - \frac{a^2 U^2}{2} \left(1-\frac{h_{+}+h_{+a}}{2}\right)\frac{\sin \frac{\bar U_h}2 \sin\frac{\bar U_{h+a}}2 }{\beta^2-\frac{h_++h_{+a}}{2}}
\end{equation}
where $h_+ \equiv h_+(t,z)$, $h_{+a} \equiv h_+(t,z+a)$ and we have written the series form of $\cos aU$. After some algebra, this is shown as
\begin{equation}
    W_{yz}= \left(1 -\frac{a^4 U^4}{8} \left(1-\frac{h_{+}+h_{+a}}{2}\right)\right)I
\end{equation}
The contribution from the $W_{xz}$ is the same as above, but the quantity in brackets in the second term proportional to $h_{+a},h_+$ have the opposite sign.
However, for the plaquette in the $x-y$ plane, we get
\begin{equation}
 W_{xy}= I- 2 \sin^2 \frac{a \bar{U}_h}{2} \sin^2 \frac{a U_{h_+}}{2} I  
\end{equation}
where $U_{h_+}= U(1+ \frac12 h_+ (t,z))$, and one gets
\begin{equation}
    W_{xy}= I - \frac{a^4 U^4}{8} \left(1- \alpha^2 (1- h_{+})\right)I
\end{equation}
where we have taken $\Phi^2_2=-\alpha^2 U^2$.

\section{Solving YM equations}{\label{app:sols}}

In this section, we will see the solutions for the equations of motion for $U(t)$ and  $\Phi_2(t)$ used in Section:\ref{sec:4}. We consider the gauge field configuration as follows:
\begin{equation}
    A^a_0 = 0, A^1_1 = A^2_2 = A^3_3 = U(t), A^2_3 = \Phi_2(t),
\end{equation}
Then, the equations of motion in Euclidean spacetime become
\begin{eqnarray}
    U'' - 2 U^3 -\frac13 U \Phi_2^2 = 0,\\
    \Phi_2''-\Phi_2 U^2 = 0.
\end{eqnarray}
Now consider the ansatz proportional to the Sine Jacobi elliptic functions as
\begin{equation}
    U(t) = c_1 {\rm sn} (ic_1 \sqrt{a} t, -1), \& \ \Phi_2(t) = c_2 {\rm sn} (ic_2 \sqrt{b} t,-1).
\end{equation}
Then, by inserting into the equations of motion, we get
\begin{eqnarray}
    2(-1+a) c_1^3 {\rm sn}^3 - \frac13 c_1 c_2^2 {\rm sn}\  \bar{\rm sn}^2 = 0,\\
    2bc_2^3 \bar{\rm sn}^3 - c_1^2 c_2 {\rm sn}^2 \bar{\rm sn} = 0,
\end{eqnarray}
where ${\rm sn} = {\rm sn} (ic_1 \sqrt{a} t, -1)$ and $\bar {\rm sn} = {\rm sn} (ic_2 \sqrt{b} t,-1)$. At finite temperature, all the fields should be in thermal equilibrium; therefore, periodic for gauge fields. This translates to the condition:
\begin{equation}{\label{appeq:equalities}}
    ic_1\sqrt{a} = ic_2 \sqrt{b} = \frac{4 K(-1)}{\beta}.
\end{equation}
Using the above first equality, one gets $sn =\bar{sn}$. Then, the simplified equations are as follows: 
\begin{equation}{\label{appeq:eq}}
    2(-1+a) c_1^3 - \frac13 c_1 c_2^2 = 0, \ \ 
    2 b c_2^3 - c_1^2 c_2 = 0.
\end{equation}
By solving the above equations, we get the relation
\begin{equation}
    a = 1 + \frac{1}{12b}.
\end{equation}
Then, using the first equality (Eq. \ref{appeq:equalities}) and the second equation of Eq.(\ref{appeq:eq}), one gets $a=1/2$ and $b=-1/6$. Thus, all the parameters are fixed now as they should be. Finally, one can rewrite the fluctuation as 
\begin{equation}
    \Phi_2(t) = \frac{i}{\sqrt{3}} U(t).
\end{equation} 

\bibliographystyle{unsrt}  
\bibliography{ref}


\end{document}